\documentclass[aip,jcp,reprint]{revtex4-2}

\usepackage[version=3]{mhchem} 
\usepackage{amsmath}
\usepackage{amssymb}
\usepackage{soul}
\usepackage{mathrsfs}
\usepackage{xcolor}
\usepackage{booktabs}
\usepackage{units}
\usepackage{bm}
\usepackage{subcaption}

\usepackage{graphicx}
\usepackage{dcolumn}

\usepackage[utf8]{inputenc}
\usepackage[T1]{fontenc}
\usepackage{mathptmx}
\usepackage{etoolbox}

\DeclareMathOperator{\Tr}{Tr}
\renewcommand{\vec}[1]{\boldsymbol{#1}} 
\newcommand{\mat}[1]{\mathbf{#1}}       
\newcommand{\ReSpect}{\textsc{ReSpect}}

\makeatletter
\def\@email#1#2{%
 \endgroup
 \patchcmd{\titleblock@produce}
  {\frontmatter@RRAPformat}
  {\frontmatter@RRAPformat{\produce@RRAP{*#1\href{mailto:#2}{#2}}}\frontmatter@RRAPformat}
  {}{}
}%
\makeatother
\begin{document}

\preprint{AIP/123-QED}

\title{Linear-Response Quantum-Electrodynamical Density Functional Theory
Based on Two-Component X2C Hamiltonians}
\author{L. Konecny}
\email{lukas.konecny@uit.no.}
\affiliation{Department of Inorganic Chemistry, Faculty of Natural Sciences, Comenius University, SK-84215 Bratislava, Slovakia}
\affiliation{Hylleraas Centre for Quantum Molecular Sciences, Department of Chemistry, UiT The Arctic University of Norway, N-9037 Troms{\o}, Norway}
\affiliation{Max Planck Institute for the Structure and Dynamics of Matter, Center for Free Electron Laser Science, Luruper Chaussee 149, 22761 Hamburg, Germany}

\author{V.P. Kosheleva}
\email{valeriia.kosheleva@mpsd.mpg.de}
\affiliation{Max Planck Institute for the Structure and Dynamics of Matter, Center for Free Electron Laser Science, Luruper Chaussee 149, 22761 Hamburg, Germany}

\author{M. Ruggenthaler}
\email{michael.ruggenthaler@mpsd.mpg.de}
\affiliation{Max Planck Institute for the Structure and Dynamics of Matter, Center for Free Electron Laser Science, Luruper Chaussee 149, 22761 Hamburg, Germany}

\author{M. Repisky}%
\email{michal.repisky@uit.no.}
\affiliation{Hylleraas Centre for Quantum Molecular Sciences, Department of Chemistry, UiT The Arctic University of Norway, N-9037 Troms{\o}, Norway}
\affiliation{Department of Physical and Theoretical Chemistry, Faculty of Natural Sciences, Comenius University, SK-84215 Bratislava, Slovakia}

\author{A. Rubio}
\email{angel.rubio@mpsd.mpg.de}
\affiliation{Max Planck Institute for the Structure and Dynamics of Matter, Center for Free Electron Laser Science, Luruper Chaussee 149, 22761 Hamburg, Germany}
\affiliation{Initiative for Computational Catalysis (ICC), The Flatiron Institute, 162 Fifth Avenue, New York, New York 10010, USA}

\date{\today}

\begin{abstract}
Linear-response quantum electrodynamical density functional theory (QEDFT) enables the description of molecular spectra under strong coupling to quantized photonic modes, such as those in optical cavities. Recently, this approach was extended to the relativistic domain using the four-component Dirac--Coulomb Hamiltonian. To provide a computationally efficient yet accurate alternative---particularly for modeling 2D spectra or collective coupling for large, heavy-element systems---this article introduces a two-component linear-response QEDFT method based on exact two-component (X2C) Hamiltonian models. We derive how the parent four-component Hamiltonian for coupled electron--photon systems undergoes the X2C transformation. Moreover, we show that, under common weak-field and dipole approximations, it suffices to apply the X2C transformation only during the ground-state self-consistent field procedure, with the subsequent calculations performed fully in the two-component regime using the same X2C decoupling matrix. The current implementation includes the atomic mean-field (amfX2C), extended atomic mean-field (eamfX2C), and molecular mean-field (mmfX2C) Hamiltonian models. Benchmark calculations demonstrate that the X2C approach closely reproduces reference four-component results, enabling us to efficiently tackle systems that would be otherwise computationally too expensive. As applications, we compute 2D spectra of a mercury porphyrin complex in a Fabry--Pérot cavity, demonstrating off-resonant coupling and the appearance of multiple polaritonic branches. We also study a chain of AuH molecules, showing that collective coupling can locally modify chemical properties of a molecule with a perturbed bond length.
\end{abstract}

\maketitle

\section{\label{sec:Introduction}Introduction}

Molecular properties can be changed by modifying the electromagnetic
field, which is the foundation of two major directions in modern chemistry: Laser driving and
cavity-materials engineering. While the former employs intense light pulses to control electrons
and nuclei in molecules, the latter is based on strong light-matter coupling that occurs in
optical cavities. Importantly, the effects of strong coupling between matter and cavity photons
also occur in the dark, i.e.\ only due to the vacuum fluctuations of the modes of the cavity without the need for external pumping that might otherwise destroy the molecules.~\cite{Garcia-Vidal2021}
Therefore, cavities represent an attractive way for engineering properties on demand, modifying
chemical reactions rates and outcomes~\cite{Hutchison2012}, transport properties~\cite{Sandik2024},
and inducing new states of matter.
These intriguing prospects create a demand for theoretical methods capable of predicting
the effects of light--matter coupling on materials placed in cavities. The strong coupling
of light and matter inside optical cavities requires a description at the level of
quantum electrodynamics (QED) that treats both light and matter as dynamical variables.
Moreover, the focus on accurately predicting the effects of light--matter coupling on the
matter subsystem forces the shift from few-level models to accurate \textit{ab initio}
methods built on advances made in quantum chemistry and materials science.~\cite{Ruggenthaler2023, Foley2023}
Various approaches combining electronic structure methods with quantum optical techniques
have been developed for this purpose and include
quantum-electrondynamical density function theory (QEDFT)~\cite{Ruggenthaler2011,Ruggenthaler2014,Tokatly2013,Ruggenthaler2015,Penz2023,Flick2019,Yang2021},
and quantum-electrondynamical versions of Hartee--Fock theory~\cite{Buchholz2019,Haugland2020, Buchholz2020},
configuration interaction~\cite{McTague2022},
and coupled cluster methods~\cite{Haugland2020, Mordovina2020, DePrince2021, Monzel2024}.

While the state of the art of cavity QED is based on non-relativistic QED, there are
many phenomena in chemistry, dubbed relativistic effects, that can only be described
by including the special theory of relativity.
This is done by describing atoms and molecules by the Dirac equation.
For many-electrons systems as are considered
in quantum chemistry, solid-state physics, and materials science, the gold standard
is represented by the four-component (4c) Dirac--Coulomb Hamiltonian that combines the
relativistic kinetic energy operator with instantaneous Coulomb interaction between
electrons and between electrons and nuclei. In addition, this Hamiltonian can be extended
to include the Breit and Gaunt terms as well.
Combined with the inclusion of transverse photons as dynamical variables,
a Dirac equation-based QEDFT presents an accurate relativistic description of
electronic structure of molecules containing any elements across the periodic
table embedded in optical cavities. As such, it includes both scalar and spin--orbit (SO)
relativistic effects variationally, allowing first-principles description of SO-driven effects
such as singlet--triplet transitions. An exemplary application is the reverse intersystem
crossing that depends on the mutual position of the excited singlet and triplet state.
This can be tuned using optical cavities, while using heavy element-containing
molecules with strong SO coupling will further enhance the process.~\cite{Kena-Cohen2007, Stranius2018, Eizner2019, Hertzog2019}
Similarly, relativistic QEDFT can in future describe SO-dependent phenomena in complex
nanostructures and solids.~\cite{Xu2014, YangSH2021}
Furthermore, the Dirac equation-based QEDFT lies between low-energy non-relativistc Pauli--Fierz-based 
QED~\cite{Spohn2004,Hiroshima2019,Ruggenthaler2023} and fully second-quantized QED of different flavours.~\cite{Ryder1996,Mandl2010,Takaesu2009}
This allows to study many open questions about fundamental aspects of light--matter interactions.~\cite{Baez2014, Saue2025}

However, fully four-component (4c) relativistic calculations are often computationally demanding due to the large size of the relevant matrices, as well as the need to evaluate costly two-electron integrals involving the small-component basis functions. The computational burden becomes even more pronounced in the context of QEDFT. For instance, to determine the usual dispersion spectra of the light--matter coupled system requires multiple computational runs at varying cavity frequencies. Another very important aspect are collective effects, where an ensemble of atoms or molecules is studied, further exacerbating the computational challenge. In free-space relativistic calculations, researchers have turned to approximate two-component (2c) Hamiltonian models as more efficient alternatives. The primary advantage of 2c methods lies in their ability to simplify the problem by discarding negative-energy states (along with the associated two-electron integrals over the small-component basis), thereby reducing the complexity of the original 4c approach by half. One such 2c Hamiltonian that has gained substantial popularity in recent years is the exact two-component (X2C) Hamiltonian, which relies solely on algebraic transformations~\cite{Heully1986,Jensen2005,Kutzelnigg2005,Ilias2007,Liu2007}.

There are several variants of the X2C Hamiltonian, each differing in the choice of the parent four-component (4c) Hamiltonian used to construct the two-component (2c) model~\cite{Jensen2005,Kutzelnigg2005,Ilias2007,Liu2007,Sikkema2009,Peng2007,Liu2009,Peng2013,Filatov2013,Konecny2016,Goings2016,Liu2018,Knecht2022,Zhang2022,Ehrman2023}. The one-electron X2C (1eX2C) model employs a purely one-electron Dirac Hamiltonian as the parent, omitting two-electron interactions from the X2C decoupling transformation~\cite{Kutzelnigg2005,Liu2007,Ilias2007}. In contrast, the molecular mean-field X2C (mmfX2C) approach performs the X2C decoupling after a fully converged 4c molecular self-consistent field (SCF) calculation~\cite{Sikkema2009}. This method is typically used in post-SCF electron correlation or property evaluations. Between the 1eX2C and mmfX2C models lie several intermediate approaches, which extend the 1eX2C model by approximately incorporating two-electron interactions~\cite{Ilias2007,Peng2007,Liu2009,Peng2013,Filatov2013,Konecny2016,Goings2016,Liu2018,Knecht2022,Zhang2022,Ehrman2023}. All of these can be viewed as extensions or refinements of earlier conceptual frameworks:
(i) element- and angular-momentum-specific screening factors in the evaluation of one-electron spin-orbit (SO) integrals~\cite{Blume1962,Blume1963,Boettger2000,Filatov2013},
(ii) a mean-field SO approach~\cite{Hess1996} forming the basis of the widely used AMFI
(Atomic Mean-Field Spin--Orbit Integral) module, and (iii) a method utilizing atomic model densities derived from Kohn--Sham density functional theory~\cite{Wullen2005}.

As part of our ongoing efforts to advance the X2C framework, we recently developed two simple, computationally efficient, and numerically accurate models: the atomic mean-field X2C (amfX2C) and the extended atomic mean-field X2C (eamfX2C)~\cite{Knecht2022}. These models build upon earlier work by Liu and Cheng~\cite{Liu2018}, incorporating complete spin-orbit and scalar-relativistic corrections from two-electron interactions, whether arising from the Coulomb, Coulomb--Gaunt, or Coulomb--Breit Hamiltonians. Moreover, they are designed to reflect the specifics of the underlying correlation framework (e.g., Kohn--Sham density functional theory), enabling the incorporation of tailored exchange--correlation corrections. Both X2C models have also been extended to property calculations using either linear response theory or real-time approaches~\cite{Konecny2023,Moitra2023}.

In this article, we introduce the X2C transformed QEDFT Hamiltonian
and linear response equations, including the picture-changed two-electron contributions
in the spirit of the atomic mean field X2C.
Section~\ref{sec:Theory} presents the theory, where in Sec.~\ref{sec:relQEFT} the 4c Hamiltonian and the
equations of motion (EOM) for the coupled light--matter system
are defined. In Sec.~\ref{sec:X2C} the X2C decoupling procedure is presented, where we first decouple the EOMs in the time domain in Sec.~\ref{sec:TD-X2C}, and then move
to the frequency domain and linear response theory in Sec.~\ref{sec:X2C-LR}. The computational details of the implementation are summarized in
Sec.~\ref{sec:CompDetails} and exemplary results are presented in Sec.~\ref{sec:Results},
first for a mercury porphyrin complex in Sec.~\ref{sec:HgP} followed by an exploration of collective
coupling in a chain of AuH molecules in Sec.~\ref{sec:Collective}. 
Firstly, we demonstrate that the method reproduces reference 4c QEDFT results at a much lower computational
cost. We then leverage this computational advantage to study collective
effects in ensembles of heavy-element atoms, and then consider 2D polaritonic spectra of large molecules.

\section{\label{sec:Theory}Theory}

\subsection{\label{sec:relQEFT}Four-component QEDFT}

\subsubsection{Four-component electron--photon Hamiltonian}
\noindent
In the long-wavelength limit and after the length-gauge transformation, a four-component (4c) relativistic Hamiltonian for atomic or molecular system coupled to the quantized modes
has the form (in SI units used throughout the article)~\cite{Ruggenthaler2014,Konecny2025}
\begin{align}
\begin{split}
  \label{eq:H_light_matter}
  \hat{H}^\mathrm{4c,QED}(t)
  =
  \hat{H}^\mathrm{4c,e}(t)
 +  \sum_{\alpha=1}^{M}
\left[\frac{\hbar}{2}\left(
\omega_\alpha \hat{q}_\alpha
- \frac{g_{\alpha} \hat{\mu}_\alpha} {\sqrt{\hbar}}
\right)^2
-\frac{\hbar }{2} 
\frac{\partial^2}{\partial \hat{q}_{\alpha}^2} \right. \\
+ \left. \sqrt{2\hbar^2 \omega_\alpha}\left[\partial_t j_{\alpha}^{\mathrm{ext}}(t)\right]
\hat{q}_{\alpha}
\right].
\end{split}
\end{align}
Here $\hat{H}^\mathrm{4c,e}(t)$ is the four-component relativistic electronic Hamiltonian. The following terms describe free photons, electron--photon interaction, and the coupling of the photonic subsystem to the (time-derivative of) an external classical current projected on the 
mode $\alpha$, $j_{\alpha}^{\mathrm{ext}}(t)$.~\cite{Konecny2025}
Each of the $M$ photon modes is described by the displacement coordinate $\hat{q}_\alpha$,
mode frequency $\omega_\alpha$, and couples to the electronic subsystem
via the molecular dipole moment projected on the mode polarization $\vec{\epsilon}_\alpha$,
$\hat{\mu}_\alpha = \vec{\epsilon}_\alpha \cdot \hat{\vec{\mu}}$.
The coupling strength $g_\alpha$ can either be determined from first-principles via solving the Maxwell's equation~\cite{Svendsen2023}, or as a phenomenological parameter that includes the effects of cavity confinement
(finite cavity volume) as well as collective-coupling effects~\cite{Ruggenthaler2023,horak2025analytic}.
Moreover, since we employ the polaritonic energy-surface partitioning~\cite{Ruggenthaler2023}, in equilibrium we can discard the nuclear contribution in the molecular dipole by re-expressing the light--matter coupling as fluctuation operators, which is equivalent to employing the zero-field condition~\cite{Schafer2020}. Alternatively, in QEDFT we can absorb the nuclear contribution as a further source term in the equation of motion (EOM) of the photonic degrees of freedom~\cite{Konecny2025}.

In the 4c relativistic case, $\hat{H}^\mathrm{4c,e}(t)$ corresponds to the 4c Dirac--Coulomb 
Hamiltonian
\begin{equation}
\begin{split}
\hat{H}^\mathrm{4c,e}(t)
=
\sum_{l=1}^N[-i\hbar c \vec{\alpha}_l \cdot\vec{\nabla}_l + \beta'_l mc^2 ] - \sum_{l=1}^N \pmb{\mathcal{E}}(t) \cdot \vec{\mu}_l \\
- \sum_{l=1}^N \sum_{k=1}^K \frac{Z_k e^2}{4 \pi \varepsilon_0} \int \frac{\rho_k(\vec{r}',\vec{R}_k) }{\left|{\vec{r}}_l-{\vec{r}'}\right|} d^3r'
+  \frac{1}{2} \sum_{l=1}^N \sum_{m \neq l}^N \frac{e^2}{4 \pi \varepsilon_0} \frac{1}{\left|{\vec{r}}_l-{\vec{r}}_m\right|}
.
\end{split}
\end{equation}
This Hamiltonian contains the free-electron Dirac Hamiltonian interacting with a
generally time-dependent external electric field $\pmb{\mathcal{E}}(t)$, as well as the longitudinal Coulomb interaction between the charged particles. 
The matter subsystem consists of $N$ electrons and $K$ nuclei. The nuclei are represented by a finite-nucleus model with a spherical Gaussian charge distribution $\rho_k$~\cite{Visscher1997}, as parametrized in Ref.~\citenum{Malkin2011}.
The electronic coordinates are
$\vec{r}_l$, the nuclear coordinates are
$\vec{R}_k$, $Z_k$ is the atomic number of the $k$-th nucleus,
$c$ the speed of the light in free-space, and
$\varepsilon_0$ the vacuum permittivity. Moreover, we employ the physical (renormalized) values of the elementary charge $e>0$ and of the electronic mass $m$.
The Hamiltonian is a 4c object due to the presence of the Dirac matrices $\vec{\alpha}$ and $\beta'$ defined as
\begin{equation}
\boldsymbol{\alpha}=
\begin{pmatrix}
0_{2}      & \sigma \\
\sigma     & 0_{2}    
\end{pmatrix}
, \quad
\beta' =
\begin{pmatrix}
0_{2} &  0_{2} \\
0_{2} & -2_{2}
\end{pmatrix}
\end{equation}
where $\vec{\sigma}$ is the vector of the $2 \times 2$ Pauli matrices that have the standard forms
\begin{equation}
\sigma_{x}=
\begin{pmatrix}
0 & 1 \\
1 & 0
\end{pmatrix}
, \quad
\sigma_{y}=
\begin{pmatrix}
0 & -i \\
i & 0
\end{pmatrix}
, \quad
\sigma_{z}=
\begin{pmatrix}
1 & 0 \\
0 & -1
\end{pmatrix}
,
\end{equation}
and $0_{2}$ and $2_{2}$ are $2 \times 2$ zero and $\mathrm{diag}(2,2)$ matrices, respectively. We note specifically that we work with the observable masses and charges of the electrons. That is, we have taken into account all (non-perturabtive and beyond dipole) contributions of the free-space vacuum and thus interpret the modes of the cavity as relative to free-space vacuum~\cite{Svendsen2023}.
A step-by-step derivation of the Hamiltonian in Eq.~\eqref{eq:H_light_matter} starting from the full QED Hamiltonian is available in Ref.~\citenum{Konecny2025}. 

The time evolution of the full 4c coupled electron--photon wave function
$\Psi$ is then dictated by $\hat{H}^\mathrm{4c,QED}(t)$ via solving

\begin{equation}
i \hbar \partial_t \Psi\left( \{\vec{r}_i\}_{i=1}^{N}; \{q_j\}_{j=1}^{M} ;t\right)= \hat{H}^\mathrm{4c,QED}(t)  \Psi\left( \{\vec{r}_i\}_{i=1}^{N}; \{q_j\}_{j=1}^{M} ;t\right)
.
\label{eq:full_e-p_EOM}
\end{equation}
In practice, approximate methods of quantum chemistry and quantum optics have to be applied
to achieve a computationally feasible solution of Eq.~\eqref{eq:full_e-p_EOM} for a system
of many electrons and photons.~\cite{Ruggenthaler2023, Foley2023}

\subsubsection{Four-component QEDFT equations of motion}
\noindent
This article employs density functional theory (DFT), specifically quantum electrodynamical DFT (QEDFT), which describes coupled electron--photon systems in terms of the (one-body) electron density and the photon displacement coordinate.~\cite{Tokatly2013,Ruggenthaler2014,Flick2019} In the Kohn--Sham (KS) DFT formalism, the electronic system is represented as an auxiliary system of independent
particles whose wave function is a Slater determinant composed of $N^\mathrm{occ}$ occupied KS molecular spin-orbitals (MOs), 
$\varphi^\mathrm{4c}_j(\vec{r},t)$.
Further we will use indices $i$, $j$ to denote occupied, $a$, $b$ virtual, and $p$, $q$ general MOs.
The electron probability density of the real system is reproduced by the density of this auxiliary system
\begin{equation}
n^\mathrm{4c}(\vec{r}, t)
= 
\sum_{j=1}^{N^\mathrm{occ}}\left|\varphi^\mathrm{4c}_{j}(\vec{r},t)\right|^2
=
\sum_{j=1}^{N^\mathrm{occ}}\varphi^\mathrm{4c\,\dagger}_{j}(\vec{r},t)\varphi^\mathrm{4c}_{j}(\vec{r},t)
.
\label{eq:density}
\end{equation}
Moreover, we will employ the the electron spin density of the auxiliary system, which is defined by
\begin{equation}
    s_k^\mathrm{4c}(\vec{r}, t)
    = 
    \sum_{j=1}^{N^\mathrm{occ}}\varphi^\mathrm{4c\,\dagger}_{j}(\vec{r},t)
    \Sigma_k \varphi^\mathrm{4c}_{j}(\vec{r},t)
    ,
   \quad
   \Sigma_{k} \equiv
   \begin{pmatrix}
       \sigma_{k} & 0_2 \\
       0_2        & \sigma_{k}
   \end{pmatrix}
.
\label{eq:SpinDensity}
\end{equation}
The KS spin-orbitals are expressed as a linear combination of 4c atomic orbital (AO) basis functions $X^\mathrm{4c}_{\mu}(\bm{r})$ as
\begin{equation}  
   \label{eq:MO_coeffs}
   \varphi^\mathrm{4c}_{p} (\vec{r},t) = X^\mathrm{4c}_{\mu}(\vec{r}) C^\mathrm{4c}_{\mu p}(t)
   ,
\end{equation}
where $C^{\mathrm{4c}}_{\mu p}$ are the MO coefficients.
Note that each $X^\mathrm{4c}_{\mu}(\vec{r})$ is a $4\times 4$ matrix, and both $\varphi^\mathrm{4c}_{p} (\vec{r})$ and $C^\mathrm{4c}_{\mu p}$ are column vectors of length four, meaning that Eq.~\eqref{eq:MO_coeffs} involves an implicit matrix--vector multiplication. A summation over repeated indices is assumed here, as well as throughout the text; however, in some cases, we explicitly write out the summation to enhance clarity. The 4c AO basis functions $X^\mathrm{4c}_{\mu}(\vec{r})$ are built using the restricted kinetic balance (RKB) prescription~\cite{Stanton1984}
\begin{equation}
   \label{eq:4Clcao}
   X^\mathrm{4c}_{\mu} (\vec{r}) =
   \begin{pmatrix}
      1_{2} & 0_{2} \\
      0_{2} & \tfrac{1}{2mc}(\vec{\sigma}\cdot\vec{p}) \\
   \end{pmatrix}
   \chi_{\mu}(\vec{r})
   ,
\end{equation}  
where $\vec{p}$ is the linear momentum operator, and the functions $\chi_{\mu}(\vec{r})$ are 
elements of a real scalar basis set of size $\mathcal{N}$, in our implementation
chosen as Gaussian-type orbitals (GTO). Since the X2C transformation is performed in an orthonormal basis, which provides better control over potential linear dependencies and simplifies the construction of the appropriate metric, the following discussion assumes that the initial GTO basis has already been orthonormalized. From now on, Greek indices $\mu$, $\nu$,
$\kappa$, $\lambda$ numerate the elements of the orthonormal AO basis.

In this basis, the 4c electron probability and spin densities are
\begin{subequations}
\begin{align}
n^{\mathrm{4c}}(\vec{r},t) 
& = 
\Omega_{0,\mu\nu}^{\mathrm{4c}}(\vec{r}) D^{\mathrm{4c}}_{\nu\mu}(t) 
, \\
s_k^{\mathrm{4c}}(\vec{r},t) 
& = 
\Omega_{k,\mu\nu}^{\mathrm{4c}}(\vec{r}) D^{\mathrm{4c}}_{\nu\mu}(t) 
,
\end{align}
\end{subequations}
where
$D^{\mathrm{4c}}_{\nu\mu}(t)$ is the 4c time-dependent density matrix
\begin{gather}
  \label{eq:D_4c}
  D_{\nu\mu}^\mathrm{4c}(t)
  =
  \sum_{j=1}^{N^\mathrm{occ}}
  C^\mathrm{4c}_{\nu j}(t)
  {C^\mathrm{4c\dagger}_{\mu j}}(t)
  .
\end{gather}
Here the overlap and spin distribution functions are defined as
\begin{subequations}
\begin{align}
\Omega^{\mathrm{4c}}_{0,\mu\nu}(\vec{r})
& = X_{\mu}^{\dagger}(\vec{r}) X_{\nu}(\vec{r}),
\\
\label{eq:omegaK}
\Omega^{\mathrm{4c}}_{k,\mu\nu}(\vec{r})
& =
X_{\mu}^{\dagger}(\vec{r}) \Sigma_k X_{\nu}(\vec{r})
.
\end{align}
\end{subequations}
In the following text we will use a joint notation for the electron densities:
$\vec{\rho}^{\mathrm{4c}}(\vec{r},t)
= \left(n^\mathrm{4c}(\vec{r},t), s_x^\mathrm{4c}(\vec{r},t), s_y^\mathrm{4c}(\vec{r},t), s_z^\mathrm{4c}(\vec{r},t)\right)$.

In an \textit{orthonormal} AO basis, the coupled 4c KS EOMs for the basic variables of the electron--photon system are
\begin{subequations}
\label{eq:ep_EOM}
\begin{align}
\label{eq:el_EOM}
i\hbar \partial_t C_{\mu j}^{\mathrm{4c}}(t,\pmb{\mathcal{E}})
& =
F^{\text{4c}}_{\mu\nu}\left(t,\pmb{\mathcal{E}}\right) C_{\nu j}^{\mathrm{4c}}(t,\pmb{\mathcal{E}})
, \\
\label{eq:ph_EOM}
\left(\partial_t^2 + \omega_\alpha^2 \right) q_\alpha (t,\pmb{\mathcal{E}}) 
& = 
-\sqrt{2}\partial_t j_{\alpha}^{\mathrm{4c}}(t,\pmb{\mathcal{E}}).
\end{align}
\end{subequations}
They are derived from Eq.~\eqref{eq:full_e-p_EOM} upon moving to the KS DFT framework and adapting to a basis set representation from operator form in Eqs.~(35) and (36) in Ref.~\citenum{Konecny2025}. 
For later reference, we write Eq.~\eqref{eq:el_EOM} 
in a compact matrix form as
\begin{equation}
\label{eq:el_EOM_mat}
i\hbar \partial_t \mat{C}^{\mathrm{4c}}(t,\pmb{\mathcal{E}})
=
\mat{F}^{\mathrm{4c}}\left(t,\pmb{\mathcal{E}}\right) \mat{C}^{\mathrm{4c}}(t,\pmb{\mathcal{E}})
.
\end{equation}
While Eq.~\eqref{eq:el_EOM} describes the time evolution of the electronic MO coefficients, Eq.~\eqref{eq:ph_EOM} governs the time evolution of the photon field via the displacement coordinate $q_{\alpha}$. The evolution of the electronic system is driven by the Fock matrix $\mat{F}$,
\textit{i.e.} the matrix representation of the KS Hamiltonian including an interaction with
an external electromagnetic field $\pmb{\mathcal{E}}(t)$. Similarly, the evolution of the
photon subsystem is driven by the total electronic 4c current $j_{\alpha}^{\mathrm{4c}}$, which is projected on the mode $\alpha$.
Both EOMs need to be treated self-consistently, as $\mat{F}^\mathrm{4c}$ and $j_{\alpha}^{\mathrm{4c}}$ are constructed from $\vec{q}$ and $\mat{D}^\mathrm{4c}$. For clarity, we have
omitted the explicit statement of these dependencies in the equations, retaining only
the dependence on time and the external field--these being the primary variables in the forthcoming perturbative treatment. Since we focus on linear-response processes, such as absorption spectroscopy, rather than direct pumping of the cavity (which would be simulated via the time-dependent external current $j_{\alpha}^{\mathrm{ext}}(t)$), we will not carry on the potential dependence on this external parameter.

Let us now provide a detailed explanation of the right hand sides of Eqs.~\eqref{eq:el_EOM} and~\eqref{eq:el_EOM_mat}.
The Fock matrix $\mat{F}^\mathrm{4c}$ in these equations has the form
\begin{equation}
\label{eq:4cFock}
\mat{F}^\mathrm{4c}(t,\pmb{\mathcal{E}})
=
\mat{F}^\mathrm{4c,e}\left[\mat{D}^\mathrm{4c}\right] +
\mat{F}^\mathrm{4c,ep}\left[\mat{D}^\mathrm{4c},\vec{q}\right]
+ \mat{F}^\mathrm{4c,ext}
\end{equation}
and consists of three parts: the purely electron part $\mat{F}^\mathrm{4c,e}$,
the electron--photon part $\mat{F}^\mathrm{4c,ep}$, and the interaction with
an external perturbing field representing a spectrometer used
to probe the atomic and molecular systems embedded in cavities.
The external field is considered in the form of a semi-classical time-dependent
electric field $\pmb{\mathcal{E}}(t)$ that couples to the electrons via the electric dipole operator
under the long-wavelength approximation as
\begin{equation}
\label{eq:F4c_ext}
\mat{F}^\mathrm{4c,ext}
=
- \mathcal{E}_u(t) \,\mat{P}_{\!u}^\mathrm{4c}
,
\end{equation}
where the matrix representation of the electric dipole operator is
\begin{equation}
\label{eq:4cElDipole}
\begin{split}
P_{u,\mu\nu}^{\mathrm{4c}} 
& =
- e\int X_{\mu}^{\dagger}(\vec{r})(r_{u}-R_{\mathrm{G},u}) X_{\nu}(\vec{r}) d^{3}r
\\ & =
- e\int (r_{u}-R_{\mathrm{G},u}) \Omega_{0,\mu\nu}(\vec{r}) d^{3}r
.
\end{split}
\end{equation}
Here, $\vec{R}_\mathrm{G}$ is the gauge origin, set in the presented
calculations to the centre of nuclear charge.
Similarly to the Cartesian component $u\in x,y,z$, let us define a projection of the vector of dipole moment matrix onto the direction of the photon polarization
\begin{equation}
\label{eq:Pprojected}
P_{\alpha,\mu\nu}^{\mathrm{4c}} \equiv \vec{\epsilon}_\alpha \cdot \vec{P}^\mathrm{4c}_{\mu\nu}
.
\end{equation}

The elements of the electronic part of the Fock matrix in the RKB basis are
\begin{align}
\begin{split}
   \label{eq:Fock_matrix_elements}
   F_{\mu\nu}^{\mathrm{4c,e}}\left[\mat{D}^\mathrm{4c}\right]
   & =
   F_{\mu\nu}^{\mathrm{4c,e}}\left[\mat{D}^\mathrm{4c}(t,\pmb{\mathcal{E}})\right] 
   \\[5pt]
   & =
   h^{\mathrm{4c}}_{\mu\nu}
   +\displaystyle
   g^{\mathrm{4c}}_{\mu\nu,\kappa\lambda}
   D^{\mathrm{4c}}_{\lambda\kappa}(t,\pmb{\mathcal{E}}) 
   \\[5pt]
   & +
   \int v^{\text{xc}}_k\!\left[\vec{\rho}^{\mathrm{4c}}(\vec{r},t,\pmb{\mathcal{E}})\right]
   \Omega_{k,\mu\nu}^{\mathrm{4c}}(\vec{r})\,d^{3}r
   ,
\end{split}
\end{align}
and correspond term-by-term to the one-electron Dirac contribution $\mat{h}^{\mathrm{4c}}$,
the two-electron contribution, and the exchange--correlation (xc) contribution,
where the index $k$ now runs over $(0,x,y,z)$.
In this work, we use non-relativistic xc potentials
with non-collinear parametrization according to Scalmani and Frisch~\cite{Frisch2012} and relativistic extension according to Cherry and co-workers~\cite{Komorovsky2019}.
The two-electron contribution is formed of generalized anti-symmetrized electron repulsion integrals (ERIs), 
\begin{align} 
\begin{split}
   \label{eq:eri}
   g^{\mathrm{4c}}_{\mu\nu,\kappa\lambda}
   & =
   I^{\mathrm{4c}}_{\mu\nu,\kappa\lambda}
   -
   \zeta I^{\mathrm{4c}}_{\mu\lambda,\kappa\nu}
   , \\
   I^{\mathrm{4c}}_{\mu\nu,\kappa\lambda}
   & \equiv
   \frac{e^{2}}{4\pi\epsilon_{0}}
   \iint
   \Omega_{0,\mu\nu}^{\mathrm{4c}}(\vec{r}_{1})
   \frac{1}{|\vec{r}_{1}-\vec{r}_{2}|}
   \Omega_{0,\kappa\lambda}^{\mathrm{4c}}(\vec{r}_{2})
   d^{3}r_{1}d^{3}r_{2}
   ,
\end{split}
\end{align}
consisting of the direct and exact exchange terms, the latter scaled by a scalar
weight factor $\zeta$ in cases when working with hybrid functionals of density functional approximations. The electron--electron xc potential can be any of the commonly used non-relativistic 
potentials and, in case of being a hybrid, also contains the factor $\zeta$.
We note that due to the inclusion of explicit exchange contributions, we are, strictly speaking, operating within the framework of generalized KS theory.

Similarly, the electron--photon part of the Fock matrix can also be written as a sum of
one-electron, two-electron, and exchange--correlation contributions
\begin{equation}
\label{eq:ep_Fock}
\begin{split}
F_{\mu\nu}^{\mathrm{4c,ep}}\left[\mat{D}^\mathrm{4c},\vec{q}\right]
=
- \sum_{\alpha=1}^{M}\sqrt{\hbar}\omega_{\alpha} g_{\alpha} q_{\alpha}(t) P_{\alpha,\mu\nu}^{\mathrm{4c}} 
\\
+ \sum_{\alpha=1}^{M} g_{\alpha}^2 P_{\alpha,\mu\nu}^{\mathrm{4c}} P_{\alpha,\kappa\lambda}^{\mathrm{4c}} D^{\mathrm{4c}}_{\lambda\kappa}(t,\pmb{\mathcal{E}})
- \sum_{\alpha=1}^{M} g_{\alpha}^2 P_{\alpha,\mu\lambda}^{\mathrm{4c}} P_{\alpha,\kappa\nu}^{\mathrm{4c}} D^{\mathrm{4c}}_{\lambda\kappa}(t,\pmb{\mathcal{E}})
\\
+ \int v^{\mathrm{xc,ep}}_k\left[\vec{\rho}^\mathrm{4c}(\vec{r},t,\pmb{\mathcal{E}}),\vec{q}\right] \Omega_{k,\mu\nu}^{\mathrm{4c}}(\vec{r}) d^{3}r
\\
+ \sum_{\alpha=1}^{M}\sqrt{\hbar}\omega_{\alpha} g_{\alpha} q_{\alpha}(t) S_{\mu\nu}^{\mathrm{4c}}
\left( \vec{\epsilon}_\alpha \cdot \sum_I^{K} e Z_I \vec{R}_I \right)
.
\end{split}
\end{equation}
The first term on the right-hand side in the above equation describes the coupling to the displacement field of the cavity. The second line describes the coupling to the polarization field of the cavity, also called the self-energy contribution. This term arises due to the partitioning of the physical electric field of the cavity into a free-photon part described by the $\hat{q}_{\alpha}$ and the bound photon part described by the $\hat{\mu}_{\alpha}$~\cite{rokaj2018light,Schafer2020,Konecny2025}. 
The third line captures the remaining contributions when compared to the exact coupled (beyond Slater-determinant) contributions, which gives rise to electron--photon xc potential ($v^{\mathrm{xc,ep}}$).
While approximations to $v^{\mathrm{xc,ep}}$ exist~\cite{Lu2024}, it is a common practice to neglect this term in QEDFT calculations.
The last term captures the effect of the (in our case static) nuclear dipole and can be
made zero by a special choice of coordinate system.

Finally, the 4c current $j_{\alpha}^{\mathrm{4c}}(t,\pmb{\mathcal{E}})$ driving the
photon subsystem in Eq.~\eqref{eq:ph_EOM} has the form
\begin{equation}
\label{eq:KScurrent}
\partial_t j_{\alpha}^{\mathrm{4c}}(t,\pmb{\mathcal{E}})
=
\partial_t j_{\alpha}^{\mathrm{M}}\left[\mat{D}^\mathrm{4c}\right]
=
\frac{\omega_\alpha g_\alpha}{\sqrt{2\hbar}} P_{\alpha,\mu\nu}^{\mathrm{4c}} D^{\mathrm{4c}}_{\nu\mu}(t,\pmb{\mathcal{E}})
.
\end{equation}
It corresponds to the mean-field current $j_{\alpha}^{\mathrm{M}}$ created
by the electrons of the coupled system.
This is the same as in the non-relativistic case, where the exchange-correlation current is zero in the case of dipole approximation and only the mean-field contribution is important.~\cite{Ruggenthaler2014, Flick2019, Konecny2025}

This finishes the recapitulation of 4c QEDFT in algebraic matrix form represented
in an orthonormal AO basis set. The 4c EOMs can be solved in real time or by perturbation theory as was done in Ref.~\citenum{Konecny2025}.
Here, on the other hand, we will transform the 4c EOMs~\eqref{eq:ep_EOM} to a two-component X2C Hamiltonian.
To do this, we transform the electronic equation~\eqref{eq:el_EOM}
by block diagonalizing the Fock matrix.
Then we perform the transformation of the electronic properties appearing
in the photonic equation~Eq.~\eqref{eq:ph_EOM}.

\subsection{\label{sec:X2C} Exact two-component (X2C) QEDFT}

\subsubsection{Central idea of X2C transformation}
\noindent
The goal of the X2C transformation is to block-diagonalize a 4c Fock matrix $\mat{F}^\mathrm{4c}$
by a unitary decoupling matrix $\mat{U}$ as
\begin{equation}
  \label{eq:decoupled_F_C}
  \mat{\tilde{F}}^\mathrm{4c}
  =
  \mat{U}^\dagger
  \mat{F}^\mathrm{4c}
  \mat{U}
  \equiv
  \begin{pmatrix}
    \mat{\tilde{F}}^\mathrm{LL} & \mat{0}_{2}                    \\[5pt]
     \mat{0}_{2}  & \mat{\tilde{F}}^\mathrm{SS}
  \end{pmatrix}
  \in \mathbb{C}^{4\mathcal{N}\times 4\mathcal{N}}
  .
\end{equation}
Here, $\mat{0}_2$ is a zero matrix of size $2\mathcal{N} \times 2\mathcal{N}$.
The tildes indicate picture-change transformed quantities,~\cite{Knecht2022}
\textit{i.e.} quantities transformed by the matrix $\mat{U}$.
The X2C transformation can be defined either for a time-independent~\cite{Kutzelnigg2005, Liu2007, Ilias2007}
or a time-dependent case.~\cite{Konecny2016, Goings2016}
Similarly, the matrix of eigenvectors of $\mat{\tilde{F}}^\mathrm{4c}$ with the original form
\begin{equation}
    \mat{C}^\mathrm{4c}
    \equiv
    \begin{pmatrix}
    \mat{C}^\mathrm{L}_\mathrm{+} & \mat{C}^\mathrm{L}_\mathrm{-} \\[5pt]
    \mat{C}^\mathrm{S}_\mathrm{+} & \mat{C}^\mathrm{S}_\mathrm{-}
    \end{pmatrix}
    ,
\end{equation}
is also transformed to a block diagonal form
\begin{gather}
  \label{eq:x4c-mos}
  \mat{\tilde{C}}^\mathrm{4c}
  =
  \mat{U}^\dagger
  \mat{C}^\mathrm{4c}
  \equiv
    \begin{pmatrix}
      \mat{\tilde{C}}^{\mathrm{L}}_{+} & \mat{0}_2                   \\[5pt]
      \mat{0}_2                    & \mat{\tilde{C}}^{\mathrm{S}}_{-}
    \end{pmatrix}
    .
\end{gather}
Here, $+$ and $-$ refer to positive and negative energy solutions, respectively.
The matrix $\mat{U}$ is conventionally parameterized as the product of a decoupling matrix and a renormalization matrix, both of which have a block structure and are defined in terms of a single matrix $\mat{R}$ with half the dimension of $\mat{U}$,~\cite{Kutzelnigg2005, Liu2007, Ilias2007}
\begin{equation}
  \mat{U} = 
  \begin{pmatrix}
      \mat{I}_2 & -\mat{R}^\dagger \\
      \mat{R}   &  \mat{I}_2
  \end{pmatrix}
  \begin{pmatrix}
      (\mat{I}_2 + \mat{R}^\dagger\mat{R})^{-1/2} & \mat{0}_2                                   \\
       \mat{0}_2                                  & (\mat{I}_2 + \mat{R}\mat{R}^\dagger)^{-1/2}
  \end{pmatrix}
  ,
\label{eq:Uparam}
\end{equation}
where $\mat{I}_2$ is a unit $2\mathcal{N} \times 2\mathcal{N}$ matrix.
The matrix $\mat{R}$ can be obtained from the one-step X2C procedure~\cite{Ilias2007}
\begin{gather}
  \label{eq:Rformula}
    \mat{C}^\mathrm{S}_\mathrm{+}
  - \mat{R} \mat{C}^\mathrm{L}_\mathrm{+}
  \overset{!}{=} \mat{0}_2,
\end{gather}
resulting from Eqs.~\eqref{eq:x4c-mos} and \eqref{eq:Uparam}.
In an orthonormal RKB basis, $\mat{R}$ scales with the speed of light as $c^{-1}$, since 
$\mat{C}^\mathrm{L}_\mathrm{+}$ and $\mat{C}^\mathrm{S}_\mathrm{+}$ are of order $c^{0}$ and $c^{-1}$,
respectively.~\cite{Konecny2023}
The X2C Fock matrix and MO coefficients are then understood as
\begin{subequations}
\label{eq:X2Cmmf-FC}
\begin{align}
   \label{eq:X2Cmmf-F}
   \mat{\tilde{F}}^{\mathrm{2c}}
   &\equiv 
   \mathbf{\tilde{F}}^\mathrm{LL}
   ,
   \\
   \label{eq:X2Cmmf-C}
   \mat{\tilde{C}}^{\mathrm{2c}}
   &\equiv 
   \mat{\tilde{C}}^{\mathrm{L}}_{+}
   .
\end{align}
\end{subequations}
A detailed description of the X2C algorithm, as implemented in our program ReSpect~\cite{Repisky2020,Repisky2025}, can be found in Ref.~\citenum{Konecny2016}.

The decoupling procedure outlined in Eqs.~\eqref{eq:decoupled_F_C}-\eqref{eq:Rformula}
was originally derived for a static case. However, in the time-dependent case of interest
in this work, the Fock matrix $\mat{\tilde{F}}^\mathrm{4c}(t)$
changes in time due its dependence on the external field $\pmb{\mathcal{E}}$, electronic density
matrix, $\mat{D}^\mathrm{4c}$, and photons via $\vec{q}$, see Eqs.~\eqref{eq:ep_EOM}.
Consequently, the MO coefficients as well the transformation matrix $\mat{U}$
inherit these dependencies.
Our goal is to show that under certain conditions, it is sufficient to find the
transformation matrix only once, either before or immediately after the
ground state SCF calculation and use that matrix in the following
time-dependent or linear response steps. This results in a significantly
simpler 2c procedure, where the block diagonalization and transformation
to a 2c form is performed only once and the rest of the calculation proceeds
in the 2c regime.

\subsubsection{\label{sec:TD-X2C} X2C Transformation of QEDFT EOMs}
\noindent
With all the dependencies in mind, let us apply the X2C transformation to the 4c electronic EOM in Eq.~\eqref{eq:el_EOM_mat}. We obtain:
\begin{equation}
\begin{split}
  \label{eq:x4c-eom}
  i \hbar \partial t \mat{\tilde{C}}^\mathrm{4c}(t,\pmb{\mathcal{E}})
  & = \mat{\tilde{F}}^\mathrm{4c}(t,\pmb{\mathcal{E}})
      \mat{\tilde{C}}^\mathrm{4c}(t,\pmb{\mathcal{E}}) 
  \\
  & +
    i \hbar \left[ \partial_t \mat{U}^\dagger(t,\pmb{\mathcal{E}}) \right]
    \mat{U}(t,\pmb{\mathcal{E}})
    \mat{\tilde{C}}^\mathrm{4c}(t,\pmb{\mathcal{E}}).
\end{split}
\end{equation}
The transformed EOM now contains an additional term involving the non-adiabatic matrix product $[\partial_t \mat{U}^\dagger(t,\pmb{\mathcal{E}})]
\mat{U}(t,\pmb{\mathcal{E}})$ with generally nonzero
off-diagonal blocks. This means, Eq.~\eqref{eq:x4c-eom} is not yet fully decoupled despite the X2C transformation.
What we aim to show next is that given small external-field perturbations, the off-diagonal blocks are negligibly small even when time- and external field-independent $\mat{U}$ is used for the decoupling.
To show this, we follow the same reasoning as in Ref.~\citenum{Konecny2023}, but here extended to the cavity-coupled case.

We begin by considering the first-order Taylor expansion of $\mat{U}$
in Eq.~\eqref{eq:Uparam} around $\| \mat{R} \| = 0$,  
\begin{equation}
  \mat{U}(t,\pmb{\mathcal{E}}) 
   =
  \begin{pmatrix}
      \mat{I}_2 & \mat{0}_2 \\
      \mat{0}_2 & \mat{I}_2
  \end{pmatrix}
  +
  \begin{pmatrix}
      \mat{0}_2 & -\mat{R}^{\dagger}(t,\pmb{\mathcal{E}}) \\
      \mat{R}(t,\pmb{\mathcal{E}})   &  \mat{0}_2
  \end{pmatrix}
  +
  O(\|\mat{R}\|^2)
  .
\label{eq:UparamTaylor}
\end{equation}
Therefore, to estimate the linear response of the matrix
$\mat{U}(t,\pmb{\mathcal{E}})$, it is sufficient to explore scaling behavior 
of the coupling matrix $\mat{R}$ with time and external electromagnetic field.
Without loss of generality, let us consider an external electromagnetic field of the harmonic time-dependent form:
$\pmb{\mathcal{E}}(t) = \lambda \mathcal{E}_0 \vec{e} \mathrm{cos}(\omega t)$. Here $\mathcal{E}_0$ is the field's unit amplitude ($\unit[1]{V/m}$) multiplied by
a dimensionless perturbation parameter $\lambda$ that controls the strength of the field. Moreover, $\vec{e}$ is the linear polarization vector of the field and $\omega$ is its frequency. A more general
field can be written as a linear combination of such monochromatic fields via the
Fourier series. The instantaneous decoupling matrix $\mat{R}$ can be expressed in a Volterra series as
\begin{equation}
  \label{eq:RTaylor}
  \mat{R}(t,\pmb{\mathcal{E}})
  = \mat{R}(0,0)
  + \pmb{\mathcal{R}}_u(\omega) \mathcal{E}_0 e_u \mathrm{cos}(\omega t) \lambda 
  + O(\lambda^2),
\end{equation}
where $\pmb{\mathcal{R}}$ is the first order correction. From Eq.~\eqref{eq:RTaylor},
the time derivative and linear response of $\mat{R}$ can be obtained as follows
\begin{subequations}
\begin{alignat}{2}
  \label{eq:Rdot}
  \partial_t \mat{R}(t,\pmb{\mathcal{E}})
  &= -\omega\, \pmb{\mathcal{R}}_u(\omega) \,\mathcal{E}_0 e_u \mathrm{sin}(\omega t) \lambda 
  + O(\lambda^2),
  \\
  \label{eq:Rprime}
  \mat{R}^{(1)}(t,\pmb{\mathcal{E}})
  &= \pmb{\mathcal{R}}_u(\omega) \, \mathcal{E}_0 e_u\mathrm{cos}(\omega t) \lambda
  .
\end{alignat}
\end{subequations}
Therefore, the time derivative and first order correction to $\mat{R}$ scale with
the relevant parameters as
\begin{subequations}
\label{eq:Rdotprime}
\begin{alignat}{2}
  \label{eq:Rdot1}
  \partial_t \mat{R}(t,\pmb{\mathcal{E}})
  &=
  O(\lambda\omega c^{-1}),
  \\
  \label{eq:Rprime1}
  \mat{R}^{(1)}(t,\pmb{\mathcal{E}})
  &=
  O(\lambda c^{-1})
  .
\end{alignat}
\end{subequations}
To obtain the dependencies in Eqs.~\eqref{eq:Rdotprime} we have used the $c^{-1}$ scaling behavior of $\mat{R}$ discussed earlier. For a molecule of size $l>\unit[1]{bohr}$, it holds that $\lambda \omega c^{-1} < \lambda \omega l c^{-1} \ll 1$, where the last inequality arises from the combination of the dipole approximation ($\omega l c^{-1} \ll 1$) and weak-field approximation ($\lambda < 1$). This implies that $\partial_t \mat{U}(t,\pmb{\mathcal{E}}) \approx 0$, allowing us to approximate $\mat{U}$ as time-independent, \textit{i.e.} $\mat{U}(t,\pmb{\mathcal{E}}) \approx \mat{U}(0,\pmb{\mathcal{E}})$. Consequently, Eq.~\eqref{eq:x4c-eom} simplifies to
\begin{equation}
  \label{eq:x4c-eom-nodU}
  i  \hbar \partial_t \mat{\tilde{C}}^\mathrm{4c} (t,\pmb{\mathcal{E}})
  = \mat{\tilde{F}}^\mathrm{4c}(t,\pmb{\mathcal{E}})
    \mat{\tilde{C}}^\mathrm{4c}(t,\pmb{\mathcal{E}}).
\end{equation}
In addition, since the relevant angular frequencies driving the electronic motion are on the order of $\unit[0.1]{au}$ and higher, we can assume that $\mat{U}^{(1)}(0,\pmb{\mathcal{E}}) \propto \lambda c^{-1} \ll 1$, which justifies
the approximation: $\mat{U}(t,\pmb{\mathcal{E}}) \approx \mat{U}(0,0)$.
Thus the X2C transformation is at all times performed by the transformation matrix
obtained in the SCF procedure that we will from now on label simply $\mat{U}$.

The use of the approximate $\mat{U}$ in Eq.~\eqref{eq:x4c-eom-nodU} leaves some off-diagonal blocks of the Fock matrix $\mat{\tilde{F}}^\mathrm{4c}(t,\pmb{\mathcal{E}})$ non-zero. Namely, the electron--photon part $\mat{\tilde{F}}^\mathrm{4c,ep}=\mat{U}^{\dagger}\mat{F}^\mathrm{4c,ep}\mat{U}$, as well as the interaction with the external field $\mat{\tilde{F}}^\mathrm{4c,ext}=\mat{U}^{\dagger}\mat{F}^\mathrm{4c,ext}\mat{U}$, as defined in 
Eq.~\eqref{eq:4cFock}.
To estimate the size of the off-diagonal terms, we will use the decoupling matrix from Eq.~\eqref{eq:UparamTaylor} with time- and external field-independent $\mat{R}$.
The off-diagonal terms have the form
\begin{align}
   \mat{\tilde{F}}^{\mathrm{LS}}
   =
   \mat{\tilde{F}}^{\mathrm{SL}^{\dagger}}
   & =
   \mat{R}^{\dagger}
   \big(
   \mat{F}^{\mathrm{SS,ep}}
   +
   \mat{F}^{\mathrm{SS,ext}}
   \big)
   -
   \big(
   \mat{F}^{\mathrm{LL,ep}}
   +
   \mat{F}^{\mathrm{LL,ext}}
   \big)
   \mat{R}^{\dagger}
   .
\end{align}
In this context, $\mat{F}^{\mathrm{LL,ep}}$ and $\mat{F}^{\mathrm{SS,ep}}$ refer to the LL and SS blocks of the electron--photon interaction matrix $\mat{F}^\mathrm{4c,ep}$, respectively. The same notation applies analogously to the external field interaction matrix $\mat{F}^\mathrm{4c,ext}$.
In practice, Eq.~\eqref{eq:x4c-eom-nodU} is solved for $\mat{\tilde{C}}^\mathrm{4c}(t,\pmb{\mathcal{E}})$ in the basis of the ground-state canonical orbitals
\begin{equation}
  \label{x4c:ansatz-mmf}
  \mat{\tilde{C}}^\mathrm{4c}(t,\pmb{\mathcal{E}})
  = 
  \mat{U}^\dagger \mat{C}^\mathrm{4c} \mat{d}(t,\pmb{\mathcal{E}})
  = 
  \begin{pmatrix}
      \mat{\tilde{C}}^{\mathrm{L}}_{+} & \mat{0}_2                   \\[5pt]
      \mat{0}_2                    & \mat{\tilde{C}}^{\mathrm{S}}_{-}
  \end{pmatrix}
   \mat{d}(t,\pmb{\mathcal{E}})
  ,
\end{equation}
where $\mat{d}$ is the matrix of unknown expansion coefficients. In this basis, 
the magnitude of the off-diagonal terms can be estimated to
\begin{align}
   \mat{\tilde{C}}^{\mathrm{L}^\dagger}_{+}
   \mat{\tilde{F}}^{\mathrm{LS}}
   \mat{\tilde{C}}^{\mathrm{S}}_{-}
   & =
   \underbrace{
   \mat{\tilde{C}}^{\mathrm{L}^\dagger}_{+}
   \mat{R}^{\dagger}
   \mat{F}^{\mathrm{SS,ep}}
   \mat{\tilde{C}}^{\mathrm{S}}_{-}
   -
   \mat{\tilde{C}}^{\mathrm{L}^\dagger}_{+}
   \mat{F}^{\mathrm{LL,ep}}
   \mat{R}^{\dagger}
   \mat{\tilde{C}}^{\mathrm{S}}_{-}
   }_{\approx O(g_{\alpha}\lambda l c^{-2})}
   \nonumber
   \\[5pt]
   & +
   \underbrace{
   \mat{\tilde{C}}^{\mathrm{L}^\dagger}_{+}
   \mat{R}^{\dagger}
   \mat{F}^{\mathrm{SS,ext}}
   \mat{\tilde{C}}^{\mathrm{S}}_{-}
   -
   \mat{\tilde{C}}^{\mathrm{L}^\dagger}_{+}
   \mat{F}^{\mathrm{LL,ext}}
   \mat{R}^{\dagger}
   \mat{\tilde{C}}^{\mathrm{S}}_{-}
   }_{\approx O(\lambda l c^{-2})}
   .
\end{align}
In these estimates, we used: $\mat{R} \propto c^{-1}$, $\mat{\tilde{C}}^{\mathrm{S}}_{-} \propto c^{-1}$,
$\tilde{P}^{\mathrm{2c}}_{\alpha,\mu\nu} \propto l$, and $q_\alpha(t) \propto \lambda$.
The last proportionality originates from the fact that all perturbations are driven by the external
field $\pmb{\mathcal{E}}(t) \propto \lambda$ that drives the molecular subsystem, which due to
the self-consistency of electron and photon EOMs, translates to the photon subsystem which then
acts back on the molecular subsystem.~\cite{Flick2019}
Now, comparing the magnitudes of the off-diagonal and diagonal elements---the latter being $c^{2}$ times larger---suggests
that the off-diagonal blocks can be neglected, allowing us to proceed by solving a fully decoupled problem.

Moreover, as customary in 2c approaches, the lower diagonal block $\tilde{\mat{C}}^\mathrm{S}_{-}(t,\pmb{\mathcal{E}})$
is disregarded as its contribution to properties investigated in this article is negligible.
Thus, we can replace Eq.~\eqref{eq:x4c-eom-nodU} with a 2c equation of the form
\begin{equation}
  \label{eq:x2c_el_eom}
  i \hbar \partial_t \mat{\tilde{C}}^\mathrm{2c}(t,\pmb{\mathcal{E}})
  = 
  \mat{\tilde{F}}^\mathrm{2c}(t,\pmb{\mathcal{E}})
  \mat{\tilde{C}}^\mathrm{2c}(t,\pmb{\mathcal{E}})
  ,
\end{equation}
where we introduced the notation
$\mat{\tilde{C}}^\mathrm{2c}(t,\pmb{\mathcal{E}}) \equiv \mat{\tilde{C}}^{\mathrm{L}}_{+}(t,\pmb{\mathcal{E}})$.
Eq.~\eqref{eq:x2c_el_eom} is again solved in the ground state MO basis either by real-time propagation~\cite{Konecny2016,Moitra2023}
or perturbation theory~\cite{Konecny2023}.
The corresponding photon equation with picture-change transformed electron-dependent terms becomes
\begin{equation}
\left(\partial_t^2 + \omega_\alpha^2 \right) q_\alpha (t) 
= 
-\sqrt{2}\partial_t \tilde{j}_{\alpha}^{\mathrm{2c}}(t)
.
\label{eq:x2c_p_EOM}
\end{equation}
In Eq.~\eqref{eq:x2c_el_eom}, the X2C-transformed Fock matrix has the form
\begin{equation}
\label{eq:2cFock}
\tilde{\mat{F}}^\mathrm{2c}(t,\pmb{\mathcal{E}})
=
\tilde{\mat{F}}^\mathrm{2c,e}\left[\tilde{\mat{D}}^\mathrm{2c}\right] +
\tilde{\mat{F}}^\mathrm{2c,ep}\left[\tilde{\mat{D}}^\mathrm{2c},\vec{q}\right]
+ \tilde{\mat{F}}^\mathrm{2c,ext}
.
\end{equation}
The forms of the individual terms in the 2c Fock matrix are:
i) the electronic term
\begin{align}
\begin{split}
   \label{eq:amfX2CFock}
   \tilde{F}_{\mu\nu}^{\mathrm{2c,e}}\left[\tilde{\mat{D}}^\mathrm{2c}\right]
   & =
   \tilde{h}^{\mathrm{2c}}_{\mu\nu}
   +
   \tilde{G}^{\text{2c}}_{\mu\nu}[\tilde{\mat{D}}^{\text{2c}}]
   +
   \tilde{V}^{\text{2c}}_{\text{xc},\mu\nu}[\tilde{\vec{\rho}}^{\text{2c}}]
   \\
   & =
   \tilde{h}^{\mathrm{2c}}_{\mu\nu}
   +\displaystyle
   \tilde{g}^{\mathrm{2c}}_{\mu\nu,\kappa\lambda}
   \tilde{D}^{\mathrm{2c}}_{\lambda\kappa}(t,\pmb{\mathcal{E}})
   \\
   & +
   \int v^{\text{xc}}_k\!\left[\tilde{\vec{\rho}}^{\mathrm{2c}}(\vec{r},t,\pmb{\mathcal{E}})\right]
   \tilde{\Omega}_{k,\mu\nu}^{\mathrm{2c}}(\vec{r})\,d^{3}r
   ,
\end{split}
\end{align}
ii) the electron--photon term
\begin{equation}
\label{eq:2c_ep_Fock}
\begin{split}
\tilde{F}_{\mu\nu}^{\mathrm{2c,ep}}\left[\tilde{\mat{D}}^\mathrm{2c},\vec{q}\right]
=
- \sum_{\alpha=1}^{M}\sqrt{\hbar}\omega_{\alpha} g_{\alpha} q_{\alpha}(t) \tilde{P}_{\alpha,\mu\nu}^{\mathrm{2c}} 
\\
+ \sum_{\alpha=1}^{M} g_{\alpha}^2 \tilde{P}_{\alpha,\mu\nu}^{\mathrm{2c}} \tilde{P}_{\alpha,\kappa\lambda}^{\mathrm{2c}} \tilde{D}^{\mathrm{2c}}_{\lambda\kappa}(t,\pmb{\mathcal{E}})
- \sum_{\alpha=1}^{M} g_{\alpha}^2 \tilde{P}_{\alpha,\mu\lambda}^{\mathrm{2c}} \tilde{P}_{\alpha,\kappa\nu}^{\mathrm{2c}} \tilde{D}^{\mathrm{2c}}_{\lambda\kappa}(t,\pmb{\mathcal{E}})
\\
+ \int v^{\mathrm{xc,ep}}_k\left[\tilde{\vec{\rho}}^\mathrm{2c}(\vec{r},t,\pmb{\mathcal{E}}),\vec{q}\right] \tilde{\Omega}_{k,\mu\nu}^{\mathrm{2c}}(\vec{r}) d^{3}r
\\
+ \sum_{\alpha=1}^{M}\sqrt{\hbar}\omega_{\alpha} g_{\alpha} q_{\alpha}(t) \tilde{S}_{\mu\nu}^{\mathrm{2c}}
\left( \vec{\epsilon}_\alpha \cdot \sum_I^{K} e Z_I \vec{R}_I \right)
,
\end{split}
\end{equation}
and finally, iii) the transformed interaction with the external field
\begin{equation}
\tilde{\mat{F}}^\mathrm{2c,ext}
=
- \mathcal{E}_u(t) \,\tilde{\mat{P}}_{\!u}^\mathrm{2c}
.
\end{equation}
In addition, the X2C transformed current in Eq.~\eqref{eq:x2c_p_EOM}
can be obtained from the corresponding 2c quantities
\begin{equation}
\label{eq:x2c_current}
\partial_t \tilde{j}_{\alpha}^{\mathrm{2c}}(t)
= \frac{\omega_\alpha g_\alpha}{\sqrt{2\hbar}} \tilde{P}_{\alpha,\mu\nu}^{\mathrm{2c}} \tilde{D}^{\mathrm{2c}}_{\nu\mu}(t).
\end{equation}

A key observation here is that all two-component Fock and current elements require the exact two-component transformed quantities, namely: the density matrix ($\mat{\tilde{D}}^{\text{2c}}$), the distribution matrices ($\tilde{\pmb{\Omega}}^{\text{2c}}_{k}$), the overlap and electric dipole integrals ($\tilde{\mat{S}}^{\text{2c}}$, $\tilde{\mat{P}}^{\text{2c}}$), as well as the one- and two-electron integrals ($\tilde{\mat{h}}^{\text{2c}}$, $\tilde{\mat{g}}^{\text{2c}}$) :
\begin{align}
\begin{split}
      \tilde{\mat{D}}^{\text{2c}}
      & =
      \Big[
      \mat{U}^{\dagger}
      \mat{D}^{\text{4c}}
      \mat{U}
      \Big]^{\text{LL}}
      ,\qquad
      \tilde{\pmb{\Omega}}^{\text{2c}}_{k}
      =
      \Big[
      \mat{U}^{\dagger}
      \pmb{\Omega}^{\text{4c}}_{k}
      \mat{U}
      \Big]^{\text{LL}}
      \\
      \tilde{\mat{S}}^{\text{2c}}
      & =
      \Big[
      \mat{U}^{\dagger}
      \mat{S}^{\text{4c}}
      \mat{U}
      \Big]^{\text{LL}}
      ,\qquad
      \tilde{\mat{P}}^{\text{2c}}
      =
      \Big[
      \mat{U}^{\dagger}
      \mat{P}^{\text{4c}}
      \mat{U}
      \Big]^{\text{LL}}
      \\      
      \mat{\tilde{h}}^{\text{2c}}
      & =
      \Big[
      \mat{U}^{\dagger}
      \mat{h}^{\text{4c}}
      \mat{U}
      \Big]^{\text{LL}}
      ,\qquad
      \mat{\tilde{g}}^{\text{2c}}
      =
      \Big[
      \mat{U}^{\dagger}
      \mat{U}^{\dagger}
      \mat{g}^{\text{4c}}
      \mat{U}
      \mat{U}
      \Big]^{\text{LL}}.
\end{split}      
\end{align}
Although the evaluation of most quantities is relatively inexpensive due to their reliance on 
two-index transformations, the X2C transformation of the 4c two-electron integrals ($\mat{g}^{\text{4c}}$) requires a costly four-index transformation, which makes two-component calculations more demanding than their four-component counterparts. Furthermore, the evaluation of the XC term necessitates the X2C transformation of both the electron densities ($\tilde{\mat{\rho}}^{\text{2c}}$) and the distribution matrices ($\tilde{\pmb{\Omega}}^{\text{2c}}_{k}$) at each grid point--once again representing a significant computational cost.

One widely used model for X2C decoupling is the molecular mean-field X2C (mmfX2C) approach, which is constructed \emph{a posteriori} from converged 4c molecular self-consistent field (SCF) solutions~\cite{Sikkema2009}. The mmfX2C model is particularly popular for post-SCF electron correlation and response theory calculations, as it exactly reproduces the positive-energy solutions of the original 4c Fock equations. However, its construction requires a full molecular 4c SCF calculation, which can be computationally demanding--especially for large systems. As a result, it is common practice to seek approaches that perform SCF iterations directly in two-component (2c) mode. In this work, in addition to mmfX2C, we extend two simpler models to QEDFT that avoid the need for a full 4c SCF solution: the atomic mean-field X2C (amfX2C) model~\cite{Knecht2022} and its extended variant, eamfX2C~\cite{Knecht2022}. 

The central idea behind our amfX2C Hamiltonian model is to construct $\tilde{\mat{G}}^{\text{2c}}$ and $\tilde{\mat{V}}^{\text{2c}}_{\text{xc}}$ in Eq.~\eqref{eq:amfX2CFock} as the sum of the two-electron (2e) and exchange-correlation (xc) contributions, using the \emph{untransformed} 2e integrals ($\mat{G}^{\text{2c}}$) and xc integrals ($\mat{V}^{\text{2c}}_{\text{xc}}$), supplemented by the picture-change correction terms $\Delta\mat{G}^{\text{2c}}$ and $\Delta\mat{V}^{\text{2c}}_{\text{xc}}$:
\begin{align}
\begin{split}
      \label{eq:G-in-amfX2C}
      \mat{\tilde{G}}^{\text{2c}}[\mat{\tilde{D}}^{\text{2c}}] 
      & = 
      \mat{G}^{\text{2c}}[\mat{\tilde{D}}^{\text{2c}}]
      + 
      \Delta\mat{G}^{\text{2c}}[\mat{\tilde{D}}^{\text{2c}}]
      ,\quad
      G^{\text{2c}}_{\mu\nu}[\mat{\tilde{D}}^{\text{2c}}]
      =
      g^{\text{2c}}_{\mu\nu,\kappa\lambda}
      \tilde{D}^{\text{2c}}_{\lambda\kappa}
      \\[5pt]
      \tilde{\mat{V}}^{\text{2c}}_{\text{xc}}\big[\tilde{\vec{\rho}}^{\text{2c}}\big]
      & =
      \mat{V}^{\text{2c}}_{\text{xc}}\big[\vec{\rho}^{\text{2c}}\big]
      +
      \Delta\mat{V}^{\text{2c}}_{\text{xc}}\big[\tilde{\vec{\rho}}^{\text{2c}}\big]
      ,\quad
      \mat{V}^{\text{2c}}_{\text{xc}}\big[\vec{\rho}^{\text{2c}}\big]
      =
      \int
      v^{\text{xc}}_{k}[\vec{\rho}^{\text{2c}}]
      \pmb{\Omega}^{\text{2c}}_{k} d^{3}r
      .
\end{split}      
\end{align}
Numerical analysis reveals that the correction terms exhibit a localized atomic character and can therefore be accurately approximated as a superposition of atomic contributions
\begin{align}
\begin{split}
      \Delta\mat{G}^{\text{2c}}
      & \simeq
      \Delta\mat{G}^{\text{2c}}_{\bigoplus}
      =
      \bigoplus_{K=1}^{\text{atoms}}
      \mat{\tilde{G}}^{\text{2c}}_{K}[\mat{\tilde{D}}^{\text{2c}}_{K}] 
      - 
      \mat{G}^{\text{2c}}_{K}[\mat{\tilde{D}}^{\text{2c}}_{K}]
      \\
      \Delta\mat{V}^{\text{2c}}_{\text{xc}}
      & \simeq
      \Delta\mat{V}^{\text{2c,xc}}_{\bigoplus}
      =
      \bigoplus_{K=1}^{\text{atoms}}
      \mat{\tilde{V}}^{\text{2c}}_{\text{xc},K}\big[\tilde{\vec{\rho}}^{\text{2c}}_{K}\big]
      -
      \mat{V}^{\text{2c}}_{\text{xc},K}\big[\vec{\rho}^{\text{2c}}_{K}\big]
      .
\end{split}      
\end{align}
Finally, the electronic Fock matrix in Eq.~\eqref{eq:amfX2CFock} is approximated within amfX2C 
as~\cite{Knecht2022}
\begin{eqnarray}
      \mat{\tilde{F}}^{\text{2c,e}}
      =
      \mat{\tilde{h}}^{\text{2c}}
      +
      \mat{G}^{\text{2c}}\big[\mat{\tilde{D}}^{\text{2c}}\big]
      +
      \Delta\mat{G}^{\text{2c}}_{\bigoplus}
      +
      \mat{V}^{\text{2c}}_{\text{xc}}\big[\vec\rho^{\text{2c}}\big]
      +
      \Delta\mat{V}^{\text{2c,xc}}_{\bigoplus}
      .
\end{eqnarray}

The main computational advantage of the amfX2C approach is that its 2e and xc picture-change (PC) corrections $\Delta{\mat{G}}^{\text{2c}}_{\bigoplus}$ and $\Delta{\mat{V}}^{\text{2c,xc}}_{\bigoplus}$ are constructed from atomic components. Consequently, all matrix elements $\Delta{G}^{\text{2c}}_{\bigoplus,\mu\nu}$ and $\Delta{V}^{\text{2c,xc}}_{\bigoplus,\mu\nu}$ are strictly zero when the basis functions $\mu$ and $\nu$ belong to different atomic centers. To account for inter-atomic PC corrections, we have also implemented within QEDFT the extended amfX2C Hamiltonian model (eamfX2C), which retains the same philosophy and form as the original amfX2C model but differs in the PC correction terms~\cite{Knecht2022}
\begin{equation}
      \mat{\tilde{F}}^{\text{2c,e}}
      =
      \mat{\tilde{h}}^{\text{2c}}
      +
      \mat{G}^{\text{2c}}[\mat{\tilde{D}}^{\text{2c}}]
      +
      \Delta\mat{G}^{\text{2c}}[\mat{\tilde{D}}^{\text{2c}}_{\bigoplus}]
      +
      \mat{V}^{\text{2c}}_{\text{xc}}[\vec\rho^{\text{2c}}]
      +
      \Delta\mat{V}^{\text{2c}}_{\text{xc}}[\tilde{\vec\rho}^{\text{2c}}_{\bigoplus}]
      .
\end{equation}
Here, $\Delta\mat{G}^{\text{2c}}$ and $\Delta\mat{V}^{\text{2c}}_{\text{xc}}$ are evaluated in a \emph{full} molecular basis using an approximate density matrix $\mat{\tilde{D}}^{\text{2c}}_{\bigoplus}$ (or a charge density $\tilde{\vec\rho}^{\text{2c}}_{\bigoplus}$) obtained by superposing the corresponding atomic terms:
\begin{align}
\begin{split}
      \Delta\mat{G}^{\text{2c}}[\mat{\tilde{D}}^{\text{2c}}]
      & \simeq
      \Delta\mat{G}^{\text{2c}}[\mat{\tilde{D}}^{\text{2c}}_{\bigoplus}]
      =
      \mat{\tilde{G}}^{\text{2c}}[\mat{\tilde{D}}^{\text{2c}}_{\bigoplus}]
      -
      \mat{G}^{\text{2c}}[\mat{\tilde{D}}^{\text{2c}}_{\bigoplus}]
      \\[5pt]
      \Delta\mat{V}^{\text{2c}}_{\text{xc}}\big[\tilde{\vec\rho}^{\text{2c}}\big]
      & \simeq
      \Delta\mat{V}^{\text{2c}}_{\text{xc}}\big[\tilde{\vec\rho}^{\text{2c}}_{\bigoplus}\big]
      =
      \mat{\tilde{V}}^{\text{2c}}_{\text{xc}}\big[\tilde{\vec\rho}^{\text{2c}}_{\bigoplus}\big]
      -
      \mat{V}^{\text{2c}}_{\text{xc}}\big[\vec\rho^{\text{2c}}_{\bigoplus}\big]
      .
\end{split}      
\end{align}
For all details regarding to (e)amfX2C models, interested readers are referred to Refs.~\citenum{Knecht2022,Konecny2023,Moitra2023,Repisky2025}.

\subsubsection{\label{sec:X2C-LR} Linear response QEDFT within the X2C framework}

Starting from the coupled electron and photon EOMs in the 2c form (Eqs.~\eqref{eq:x2c_el_eom} and \eqref{eq:x2c_p_EOM}, respectively) we can proceed in the same steps as in the 4c regime detailed in
Ref.~\citenum{Konecny2025}.
However, since our original 4c article presents the derivation in the language
of molecular orbitals as functions, let us recast the crucial steps and expressions
in the form of MO coefficients that is more aligned with the fully algebraic
nature of the X2C method.
Expressing the cosine of the external-field perturbation as defined above in terms of exponentials, i.e.
\begin{align}
  \label{eq:Et}
  \pmb{\mathcal{E}}(t) = \lambda \mathcal{E}_{0} \mathcal{F}(t) \pmb{e},
  \quad
  \mathcal{F}(t) = \frac{1}{2} \left( e^{-i\omega t} + e^{i\omega t}\right)
  ,
\end{align}
the coupled electron--photon system can be expressed in a perturbation series of the form
\begin{align} 
  \label{eq:C2c-PTexpansion}
  \tilde{C}^\mathrm{2c}_{\mu i}(t,\pmb{\mathcal{E}})
  & = 
  \tilde{C}^\mathrm{2c}_{\mu p} 
  \left[
     \delta_{pi}
     +
     \lambda d_{pi}^{(1)}(t)
     +
     \lambda^2 d_{pi}^{(2)}(t)
     + \ldots
  \right]
  e^{-i\varepsilon_{i}t}
  .
\end{align}
The zeroth-order solutions are exactly the reference MOs, hence the Kronecker delta, that without an external field evolve in time only in phase with the factor $\exp(-i\varepsilon_{i}t)$. The first order term was expressed via $\mat{d}^{(1)}$-coefficients that express the correction to occupied time-dependent MOs
as a linear combination of ground state MOs.
Moreover, the perturbation expansion for the photon coordinate has the form
\begin{equation}
\label{eq:q-PTexpansion}
    q_\alpha (t) = q^{(0)}_\alpha (t) + \lambda q^{(1)}_\alpha (t) + \lambda^2 q^{(2)}_\alpha (t) + \ldots \;.
\end{equation}
Since we expressed the Fock matrix defined in Eqs.~\eqref{eq:2cFock}, \eqref{eq:amfX2CFock}, and
\eqref{eq:2c_ep_Fock} as well as the current defined in Eq.~\eqref{eq:x2c_current} in terms
of the electron density matrix, we define its first order correction as
\begin{equation}
\label{eq:Dmat1}
\tilde{D}^{\mathrm{2c}\,(1)}_{\nu\mu}(t)
\equiv
\tilde{C}^\mathrm{2c}_{\nu a}\tilde{C}^{\mathrm{2c}^\dagger}_{\mu i} d_{ai}^{(1)}(t)
+ \tilde{C}^\mathrm{2c}_{\nu i}\tilde{C}^{\mathrm{2c}^\dagger}_{\mu a} d^{(1)^*}_{ai}(t)
.
\end{equation}
Here we took into account that to first order only the virtual--occupied
and occupied--virtual blocks of the $\mat{d}^{(1)}$ coefficients are non-zero and complex
conjugate with each other. This is a consequence of the normalization condition for the MOs.~\cite{Konecny2019}
Furthermore, the first-order expansion coefficients are parametrized as
\begin{equation}
  \label{eq:rspXYansatz}
  \mat{d}^{(1)}(t)
  = 
  \mat{X} e^{- i \omega t} + \mat{Y}^\ast e^{i \omega t}
  ,
\end{equation}
where the time dependence comes from the driving external field.
The expression in Eq.~\eqref{eq:rspXYansatz} is a step in the solution of a differential
equation by the method of undetermined coefficients.

The perturbation series in Eqs.~\eqref{eq:C2c-PTexpansion} and \eqref{eq:q-PTexpansion} are
inserted into the EOMs in Eqs.~\eqref{eq:x2c_el_eom} and \eqref{eq:x2c_p_EOM}.
The first-order equations capture the linear contributions to excitations in the coupled
electron--photon system. The equations for the first order contributions $\mat{d}^{(1)}$
and $\vec{q}^{(1)}$ are again coupled differential equations.
Specifically, the equation for the first order correction to electron MO coefficients has the form
\begin{align}
\begin{split}
    \label{eq:d1CoefEOM}
    i \partial_t d^{(1)}_{ai}(t)
    = &
    A^\mathrm{2c}_{ai,bj} (\omega)\, d^{(1)}_{bj}(t) + B^\mathrm{2c}_{ai,bj} (\omega) \, d^{(1)*}_{bj}(t) \\
    & + \frac{g_\alpha^2}{2} \tilde{P}_{\alpha,ai}^{\mathrm{2c}} \tilde{P}_{\alpha,jb}^{\mathrm{2c}}\, d^{(1)}_{bj}(t)
    + \frac{g_\alpha^2}{2} \tilde{P}_{\alpha,ai}^{\mathrm{2c}} \tilde{P}_{\alpha,bj}^{\mathrm{2c}}\, d^{(1)*}_{bj}(t) \\
    & + \sqrt{\hbar} \omega_\alpha g_\alpha \tilde{P}_{\alpha,ai}^{\mathrm{2c}} q^{(1)}_\alpha (t)
    + \tilde{P}^\mathrm{2c}_\mathrm{ai} e^{-i \omega t} + \tilde{P}_\mathrm{ai}^{\mathrm{2c}^*} e^{i \omega t}
    ,
\end{split}
\end{align}
where the matrices $\mat{A}^\mathrm{2c}$ and $\mat{B}^\mathrm{2c}$ are
representations of the response kernel in the canonical MO basis
\begin{subequations}
\label{eq:ABkernel}
\begin{align}
  \label{eq:Aterm}
  A^\mathrm{2c}_{ai,bj} 
  &= 
  \omega_{ai}\delta_{ab}\delta_{ij} 
  +
  \left( g^\mathrm{2c}_{\mu\nu,\kappa\lambda} + k^\mathrm{xc}_{\mu\nu,\kappa\lambda} \right)
  \tilde{C}^{\mathrm{2c}\dagger}_{\mu a} \tilde{C}^\mathrm{2c}_{\nu i}
  \tilde{C}^{\mathrm{2c}\dagger}_{\kappa j} \tilde{C}^\mathrm{2c}_{\lambda b}
  ,
  \\
  \label{eq:Bterm}
  B^\mathrm{2c}_{ai,bj} 
  &=
  \left( g^\mathrm{2c}_{\mu\nu,\kappa\lambda} + k^\mathrm{xc}_{\mu\nu,\kappa\lambda} \right)
  \tilde{C}^{\mathrm{2c}\dagger}_{\mu a} \tilde{C}^\mathrm{2c}_{\nu i}
  \tilde{C}^{\mathrm{2c}\dagger}_{\kappa b} \tilde{C}^\mathrm{2c}_{\lambda j}
  ,
  \\
  \label{eq:Kterm}
  \mathbf{k}^\mathrm{xc} &= \mathbf{k}^\mathrm{xc}
       \Big( \boldsymbol{\Omega}^\mathrm{2c}, \mathbf{\tilde{D}}^\mathrm{2c} \Big)
  .
\end{align}
\end{subequations}
Here we additionally assumed the adiabatic approximation to the DFT kernel by making it
independent on $\omega$ as well as neglected the photon--electron kernel (i.e. assumed
the photon random phase approximation)~\cite{Flick2019}.
Furthermore, we discarded the exchange contribution to the self energy which has a precedence
in linear response QEDFT.~\cite{Flick2019, Yang2021}
The term responsible for the coupling to the nuclear dipole moment is additionally zero in linear response equations within the clamped nuclei approximation. This can be seen
from the fact that this term in the Fock matrix in Eq.~\eqref{eq:2c_ep_Fock} is proportional
to the overlap matrix $\tilde{S}^\mathrm{2c}_{\mu\nu}$, which will in the
linear response regime in the canonical MO basis translate to $\tilde{S}^\mathrm{2c}_{ai} = 0$
(equivalently in 4c Eq.~\eqref{eq:ep_Fock}, $S^\mathrm{4c}_{\mu\nu}$ would lead to $S^\mathrm{4c}_{ai} = 0$).

The EOM for the first order correction to the photon displacement coordinate reads
\begin{equation}
    \label{eq:1stOrderMaxwell}
    \left(\partial_t^2 + \omega_\alpha^2 \right) q^{(1)}_\alpha (t)
    =
    -\sqrt{2} \partial_t \tilde{j}^{2c,(1)}_{\alpha}(t)
    .
\end{equation}
Here the first-order contribution to the current is obtained form the picture-changed KS current
defined in Eq.~\eqref{eq:x2c_current} by inserting the expression for the first-order density matrix
from Eq.~\eqref{eq:Dmat1} to yield
\begin{equation}
\begin{split}
    \partial_t \tilde{j}^{2c,(1)}_{\alpha}(t)
    & = \frac{\omega_\alpha g_\alpha}{\sqrt{2\hbar}}
    \Tr \left[ \tilde{P}_{\alpha,\mu\nu}^{\mathrm{2c}} \tilde{D}^{\mathrm{2c}\,(1)}_{\nu\mu}(t)\right]
    \\
    & =
    \frac{\omega_\alpha g_\alpha}{\sqrt{2\hbar}}
    \Tr \left\lbrace \tilde{P}_{\alpha,\mu\nu}^{\mathrm{2c}}
    \left[ \tilde{C}^\mathrm{2c}_{\nu b}\tilde{C}^{\mathrm{2c}^\dagger}_{\mu j} d_{bj}^{(1)}(t)
    + \tilde{C}^\mathrm{2c}_{\nu j}\tilde{C}^{\mathrm{2c}^\dagger}_{\mu b} d^{(1)^*}_{bj}(t) \right]
    \right\rbrace
    \\
    & =
    \frac{\omega_\alpha g_\alpha}{\sqrt{2\hbar}}
    \left[ \tilde{P}_{\alpha,jb}^{\mathrm{2c}} d_{bj}^{(1)}(t) + \tilde{P}_{\alpha,bj}^{\mathrm{2c}} d^{(1)^*}_{bj}(t) \right]
    .
\end{split}
\end{equation}
The trace in the first and second rows is over the internal multi-component structure.
The photonic EOM with electronic variables expressed at the X2C level of theory thus reads
\begin{equation}
    \label{eq:q2ndOrderDE}
    \left(\partial_t^2 + \omega_\alpha^2 \right) q^{(1)}_\alpha (t)
    =
    \frac{\omega_\alpha g_\alpha}{\sqrt{2\hbar}}
    \left[ \tilde{P}_{\alpha,jb}^{\mathrm{2c}} d_{bj}^{(1)}(t) + \tilde{P}_{\alpha,bj}^{\mathrm{2c}} d^{(1)^*}_{bj}(t) \right]
    .
\end{equation}
The differential equation for electrons in Eq.~\eqref{eq:d1CoefEOM} (as well as an analogous equation for $\mat{d}^{*}$)
and the differential equation for photons in Eq.~\eqref{eq:q2ndOrderDE} can be solved together.
Although Eq.~\eqref{eq:q2ndOrderDE} is a second-order differential equation \eqref{eq:q2ndOrderDE},
it can be transformed into two coupled first-order equations by defining the conjugate displacement ``momentum''
\begin{equation}
\label{eq:PhotonMomentum}
p_\alpha^{(1)} = -i \partial_t q_\alpha^{(1)}    
\end{equation}
and replacing in Eq.~\eqref{eq:q2ndOrderDE}
$\partial_t^2 q_\alpha^{(1)}$ with $i \partial_t p_\alpha^{(1)}$.
This results in a set of four coupled first-order equations that can be solved by utilizing
the ansatz in Eq.~\eqref{eq:rspXYansatz} and an analogous parametrization for photons,
$q^{(1)}_\alpha = \sqrt{2/\omega_\alpha} (M_\alpha - N_\alpha)$ and
$p^{(1)}_\alpha = \sqrt{2 \omega_\alpha} (M_\alpha + N_\alpha)$
as discussed in Ref.~\citenum{Konecny2025}.
The result of the derivation is an algebraic equation in the form of a generalized eigenvalue equation,
where eigenvectors contain the undetermined coefficients
$(\mat{X}_n\, \mat{Y}_n\, \mat{M}_n\, \mat{N}_n)^\mathrm{T}$,
and the eigenvalue is an excitation energy of the coupled electron--photon system $\omega_n$
with the index $n$ labeling excited states.
The equation looks as follows
\begin{widetext}
\begin{equation}
\label{eq:CavityCasidaSym}
    \begin{pmatrix}
    \mat{A}^\mathrm{2c}       + \pmb{\Delta}^\mathrm{2c}  &   \mat{B}^\mathrm{2c}       + \pmb{\Delta}^{\prime\,\mathrm{2c}}  &   \pmb{\Gamma}^\mathrm{2c}    &   \pmb{\Gamma}^\mathrm{2c}    \\
    \mat{B}^{\mathrm{2c}^\ast} + \pmb{\Delta}^{\prime\,\mathrm{2c}^\ast} &  \mat{A}^{\mathrm{2c}^\ast} + \pmb{\Delta}^{\mathrm{2c}^\ast}  &   \pmb{\Gamma}^{\mathrm{2c}^\ast}  &   \pmb{\Gamma}^{\mathrm{2c}^\ast}  \\
    \pmb{\Gamma}^{\prime\,\mathrm{2c}^\ast}                   &  \pmb{\Gamma}^{\prime\,\mathrm{2c}}              & \bm{\omega}  &   \mat{0}    \\
    \pmb{\Gamma}^{\prime\,\mathrm{2c}^\ast}                   &  \pmb{\Gamma}^{\prime\,\mathrm{2c}}              & \mat{0}      &  \bm{\omega} \\
    \end{pmatrix}
\begin{pmatrix}
\mat{X}_n \\
\mat{Y}_n \\
\mat{M}_n \\
\mat{N}_n \\
\end{pmatrix}
=
\omega_n
\begin{pmatrix}
\mat{1} &  \mat{0} & \mat{0} &  \mat{0} \\
\mat{0} & -\mat{1} & \mat{0} &  \mat{0} \\
\mat{0} &  \mat{0} & \mat{1} &  \mat{0} \\
\mat{0} &  \mat{0} & \mat{0} & -\mat{1} \\
\end{pmatrix}
\begin{pmatrix}
\mat{X}_n \\
\mat{Y}_n \\
\mat{M}_n \\
\mat{N}_n \\
\end{pmatrix}
,
\end{equation}
\end{widetext}
and resembles the working equation of linear response time dependent density functional theory
(TDDFT)~\cite{Casida1995, Casida2009, Ullrich} extended by blocks describing the photons,
and coupling terms between the particles. Note that in Ref.~\citenum{Konecny2025} we derived
a non-symmetric form of this equation, but we showed that this equation can be brought into the here shown symmetric form.
In this symmetrized version of the equation, the electron--photon and
photon--electron coupling blocks have the same form
\begin{equation}
    \Gamma^\mathrm{2c}_{ai,\alpha} = \Gamma^\mathrm{\prime\,2c}_{\alpha,ai} =
    \sqrt{\frac{\omega_\alpha}{2}}\, g_\alpha \, \tilde{P}_{\alpha,ai}^{\mathrm{2c}}
    .
\end{equation}
Moreover, in Eq.~\eqref{eq:CavityCasidaSym}, the terms $\mat{A}^\mathrm{2c}$ and $\mat{B}^\mathrm{2c}$ are defined
in Eqs.~\eqref{eq:ABkernel} while the self-energy terms are
\begin{subequations}
\label{eq:SelfEnergy}
\begin{align}
    \Delta_{ai,bj}^\mathrm{2c}  & = \frac{g_\alpha^2}{2} \tilde{P}_{\alpha,ai}^{\mathrm{2c}} \tilde{P}_{\alpha,jb}^{\mathrm{2c}} \\
    \Delta^{\prime\,\mathrm{2c}}_{ai,bj} & = \frac{g_\alpha^2}{2} \tilde{P}_{\alpha,ai}^{\mathrm{2c}} \tilde{P}_{\alpha,bj}^{\mathrm{2c}}
    .
\end{align}
\end{subequations}
Eq.~\eqref{eq:CavityCasidaSym} is expressed in terms of vectors and matrices despite the fact that
the transition vectors $\mathbf{X}_n$, $\mathbf{Y}_n$ are originally defined as matrices,
the terms $\mat{A}$, $\mat{B}$, $\pmb{\Delta}^{\mathrm{2c}}$, $\pmb{\Delta}^{\prime\,\mathrm{2c}}$ as rank 4 tensors,
and the terms $\pmb{\Gamma}^\mathrm{2c}$, $\pmb{\Gamma}^{\prime\,\mathrm{2c}}$ as rank 3 tensors, respectively.
This was achieved by considering the virtual--occupied pair indices $ai$ and $bj$ to label a single dimension.

Importantly, the 2c linear response equation~\eqref{eq:CavityCasidaSym} has the same
overall form as the 4c equation allowing the same solution strategies. While direct
solution based on inversion or elimination methods are possible for smaller systems, they
are prohibitive for larger ones due to memory and time constraints. We employ a version of
the Davidson--Olsen solver~\cite{Davidson1975, Olsen1990} that solves Eq.~\eqref{eq:CavityCasidaSym}
iteratively by projecting it into a subspace of trial vectors expanded in the course of
the iterations and solving the much smaller projected equation for the expansion coefficients.
Moreover, the solver uses a paired structure of the trial vectors first introduced for
the linear-response TDDFT equation~\cite{Komorovsky2019} extended to the photon part
of the vectors~\cite{Konecny2025} to ensure robust convergence.

\subsection{Calculation of spectra}
\label{sec:Spectrum}

The frequency-dependent polarizability~\cite{Norman2018} that defines the absorption
spectrum in QEDFT has the form
\begin{equation}
\label{eq:PolarizabilityFinalSym}
  \alpha_{uv}(\omega)
  =
  \sum_n
  \left[
  \frac{t^*_{n,u} t_{n,v}}{\omega_n+\omega+i\gamma}
  -
  \frac{t_{n,u} t^*_{n,v}}{\omega-\omega_n+i\gamma}
  \right]
  ,
\end{equation}
where the index $n$ runs over positive energy eigenvalues and the transition dipole moments
between the ground and the $n$-the excited state are
\begin{equation}
\label{eq:transitionDipoleEV}
t_{n,u}
=
X_{ai,n}(\omega) \tilde{P}^{\mathrm{2c}}_{ia,u} + Y_{ai,n}(\omega) \tilde{P}^{\mathrm{2c}}_{ai,u}
.
\end{equation}
The expression in Eq.~\eqref{eq:transitionDipoleEV} involves only the electronic variables from the
transition vector due to the fact that the dipole moment operator acts only in the electronic Fock space.
The broadening parameter $\gamma$ has been added
to transform line spectra into band spectra and should be understood only  as a convenience used when
representing a cavity by a single or few modes. Physical radiative broadening can be described
in linear-response QEDFT by considering coupling to a dense spectrum of photon modes~\cite{Flick2019}.
The absorption spectrum of a molecule in a cavity is then calculated using the usual formula for the dipole strength function
\begin{equation}
\label{eq:rspFunction}
S(\omega)
=
\frac{4\pi\omega}{3 c} \Im\, \textrm{Tr} \left[ \boldsymbol{\alpha}(\omega) \right]
,
\end{equation}
where $\Im$ denotes the imaginary part, and the trace Tr is performed over the Cartesian components. As in our previous work and as is common in the field of polaritonic chemistry, we used the same formula for the spectral
function as is used in the free space. We have thus neglected subtleties of performing a proper rotational averaging for the molecule-in-cavity set-up.~\cite{Sidler2020,Schnappinger2023}

\section{\label{sec:CompDetails} Computational Details}

All the presented calculations were performed using the relativistic spectroscopy DFT program \ReSpect{}.~\cite{Repisky2020, Repisky2025}
The linear response QEDFT calculations started from a ground state SCF reference without coupling to photons
(i.e. no ultrastrong coupling regime). The calculations were performed at the amfX2C relativistic level of
theory with already the ground state SCF done in the two-component regime. The 2c molecular
spinorbitals were constructed as linear combinations of Gaussian-type orbital (GTO) basis sets. In the
initial 4c one-electron Dirac Hamiltonian as well as atomic calculations needed for the two-electron
picture change corrections of amfX2C, the 4c atomic basis functions were created considering the restricted
kinetically balanced (RKB) relation for the small component.~\cite{Stanton1984}
In the exemplary calculations, the scalar basis sets were i) for mercury porphyrin the uncontracted Dyall's VDZ
basis sets for mercury~\cite{Dyall2010-5d} and the uncontracted Dunning's cc-pVDZ basis sets~\cite{Dunning1989}
for light elements; and ii) for AuH the uncontracted Dyall's VDZ for both gold and hydrogen.~\cite{Dyall2023-database1, Dyall2023-database2}
The density functional calculations were performed using the hybrid B3LYP~\cite{Slater1951, Vosko1980, Becke1988, Lee1988, Stephens1994}
XC approximation for the potential as well as the kernel constructed in a non-collinear fashion.
The adiabatic approximation for the XC kernel was considered in the linear response calculations.
The XC potential and kernel were evaluated using numerical integration on an adaptive molecular grid of medium
size (program default).
The linear response QEDFT calculations were performed considering the photon random phase approximation (pRPA)
that neglects the electron--photon XC potential and kernel.
Atomic nuclei were described as a finite size Gaussian charge distribution.

\section{\label{sec:Results} Results}

To showcase the amfX2C Hamiltonian model-based linear response QEDFT we consider two cases
where computational efficiency is crucial: i) a calculation of 2D spectra of a cavity-embedded
heavy-element complex, and ii) a chain of heavy element-containing molecules to study
collective coupling effects.

\subsection{\label{sec:HgP} 2D spectra of mercury porphyrin}

As the first system to demonstrate the new calculation modes enabled by X2C-accelerated linear response QEDFT,
we consider a mercury porphyrin complex. We previously investigated its cavity-modified spectra in our 4c work~\cite{Konecny2025},
where we focused on a few calculations with specific cavity parameters.
The molecular geometry and reference 4c results were taken from Ref.~\citenum{Konecny2025}.
We also note that the modification of porphyrin properties
in optical cavities is an ongoing area of research, making this class of compounds particularly interesting from the perspective
of polaritonic chemistry.~\cite{Kena-Cohen2007, Avramenko2022, Sun2022} In the presented QEDFT calculations, the molecule is positioned
in the $yz$ plane. Note, however, that the molecule is not perfectly planar, exhibiting a slight bulging in the $x$ direction, and thus
has a C4 symmetry axis coinciding with the $x$ axis. As the source of photon modes, we consider an ideal Fabry--P{\'e}rot cavity with a single
mode polarized along the $z$ axis.

The spectrum of mercury porphyrin is dominated by three main bands---labeled B, N, and L---located at 3.39, 3.76, and \unit[4.12]{eV},
respectively. However, despite the seemingly simple appearance of the spectrum, the region dominated by these bands contains many
excited states. This complexity arises from the presence of numerous low-intensity or dark states, as well as the fact that all
excited states are doubly degenerate, with each pair of bright degenerate states exhibiting mutually perpendicular transition dipole
moments. Specifically, the L peak corresponds to excitations number 81 and 82. Note that, in a cavity, only the excitation with a transition
dipole moment aligned with the cavity mode polarization couples to the photon mode. The other excitation remains uncoupled, preserving
a spectral signal at the same frequency as in the absence of the cavity. A consequence of this dense spectrum is that approximating the
molecule with a simplified model---such as a four-level system composed of the ground state and three excited states---requires a reference
ab initio calculation, such as TDDFT, to obtain at least 82 excitation energies and transition dipole moments for proper parametrization.
This requirement diminishes the practical advantages of the model-based approach, since a linear response QEDFT calculation is comparably
demanding while also accounting for self-energy terms and coupling between excited states.~\cite{Konecny2025}

Moreover, due to the proximity of the B, N, and L bands---all lying within \unit[1.0]{eV} of each other---the cavity, when in resonance
with one of these lines, also couples off-resonantly to the others. This results in more than two excited states acquiring polaritonic
character. Furthermore, due to the strong light--matter coupling and the self-consistent nature of the interaction between matter and
cavity modes, the refractive index inside the cavity changes, shifting the resonance to a different energy than the excitation energy
of the free molecule. This effect, which is due to the self-consistent change of refractive index when the cavity couples to matter, has to be taken into account also in experiments. For instance, once a Fabry-P\'erot cavity is filled with molecules, the distance between the mirrors has to be re-adjusted to be exactly on resonance. In non-self-consistent calculations, this shift must either be
estimated ad hoc or deduced from experiment.

\begin{figure}
\centering
\caption{Absorption spectra of a mercury porphyrin complex in optical cavities
with different coupling strengths $g_\alpha$ (in au), calculated using the amfX2C Hamiltonian 
(2c) and the reference four-component Hamiltonian (4c).
}
  \begin{subfigure}{0.49\textwidth}
    \includegraphics[width=\textwidth]{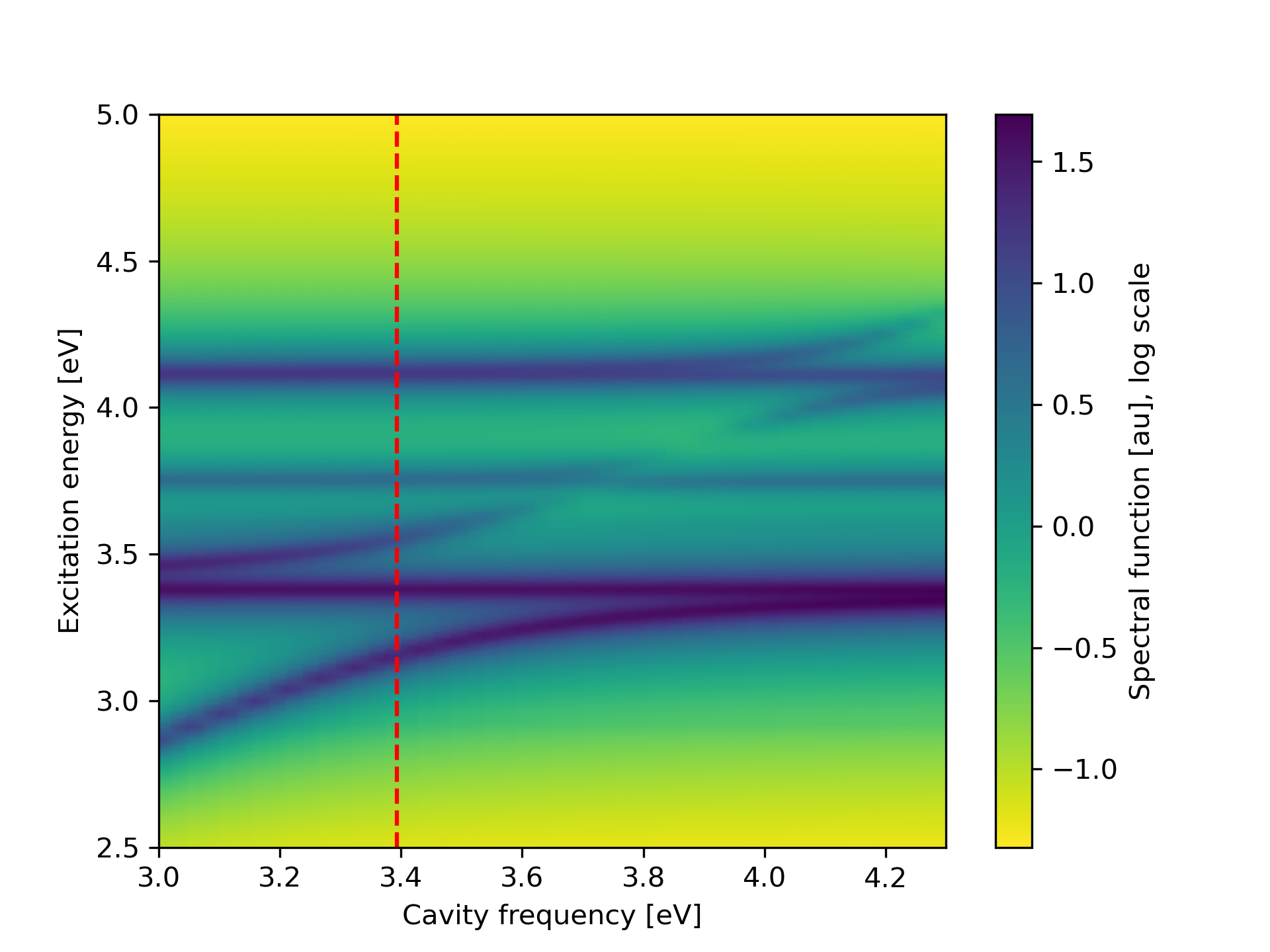}
    \caption{Calculated (2c) 2D spectrum with $g_\alpha = 0.01$.}
    \label{fig:HgP_2D_g01}
  \end{subfigure}
  \begin{subfigure}{0.49\textwidth}
    \includegraphics[width=\textwidth]{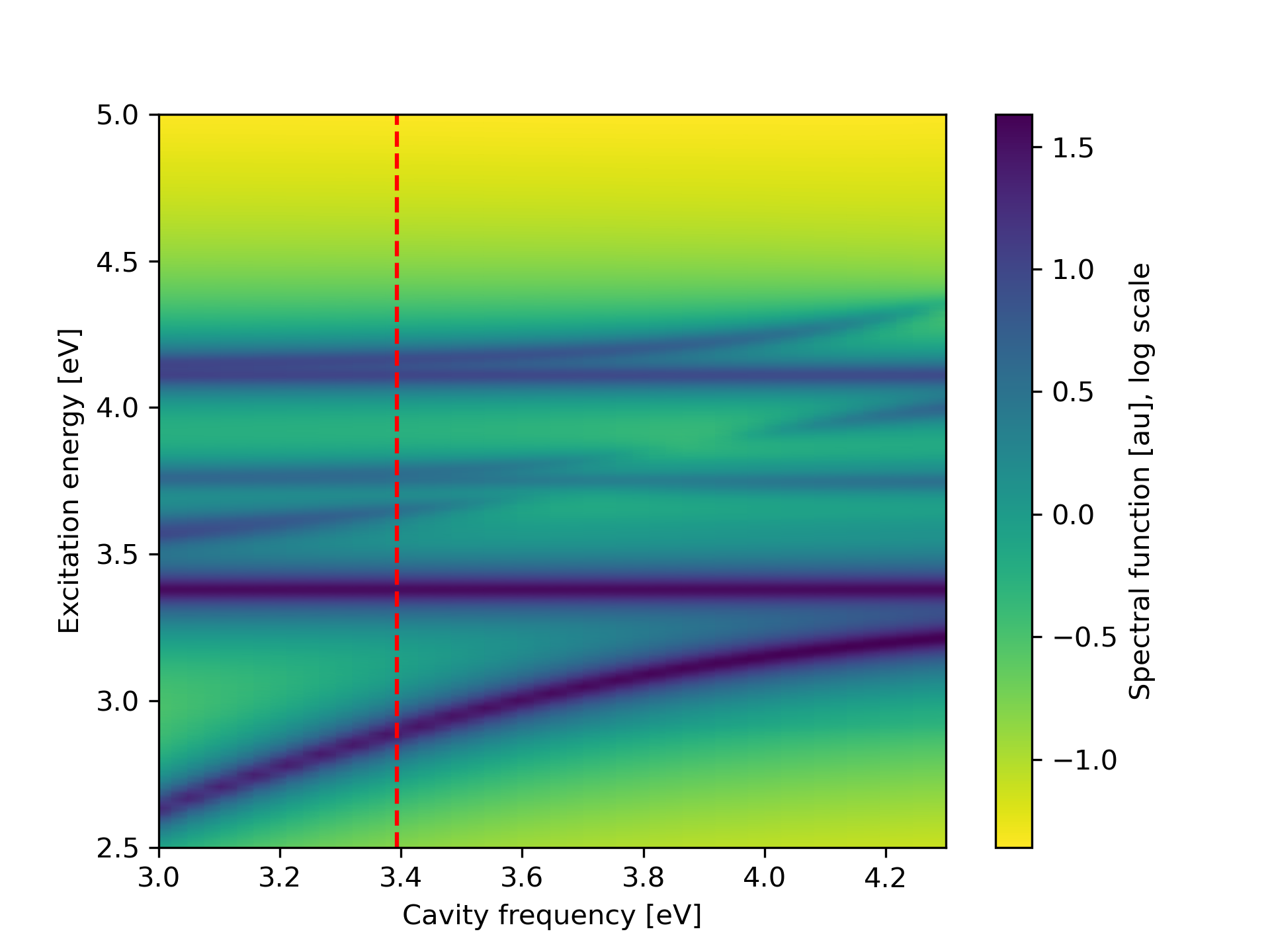}
    \caption{Calculated (2c) 2D spectrum with $g_\alpha = 0.02$.}
    \label{fig:HgP_2D_g02}
  \end{subfigure}
  \begin{subfigure}{0.49\textwidth}
    \includegraphics[width=\textwidth]{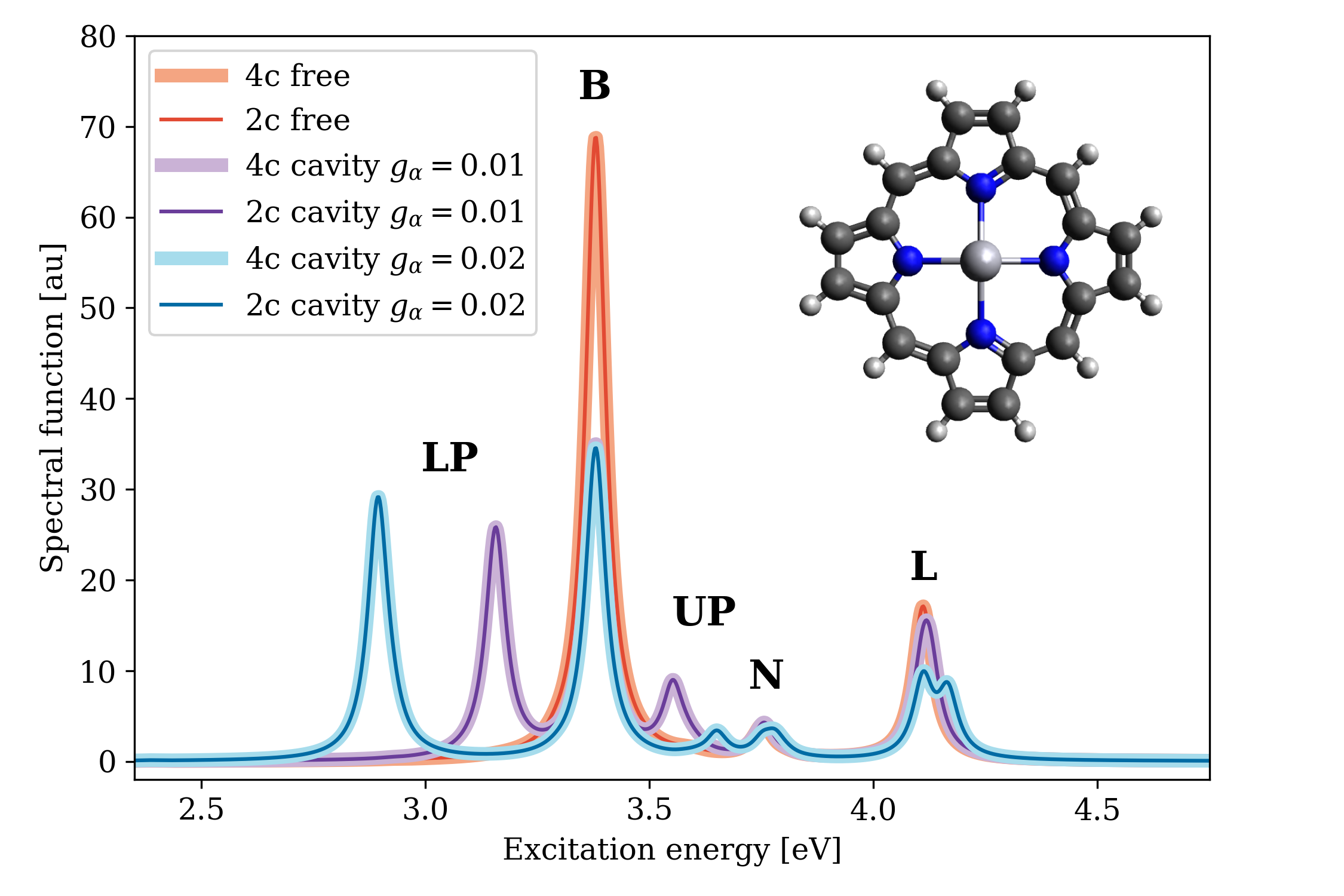}
    \caption{Spectra of a free molecule and in cavities in resonance with the B line
    [red dashed lines in panels (a) and (b)].}
    \label{fig:HgP_res}
  \end{subfigure}
  \label{fig:HgP_in_cavity}
\end{figure}

An alternative approach is to compute 2D spectra by varying the cavity frequency over a defined spectral range. To this end, we scan the
cavity mode frequency between 3 and \unit[4.35]{eV}, using coupling strengths of 0.01 and \unit[0.02]{au}, and apply an empirical
broadening parameter of $\gamma = \unit[0.027]{eV}$ to convert line spectra into band spectra. The resulting 2D spectra are shown
in Figs.~\ref{fig:HgP_2D_g01} and \ref{fig:HgP_2D_g02}. We also present 1D spectra for the free molecule and for the molecule in a cavity
tuned to resonance with the free-molecule B line, indicated by a red dashed line in the 2D spectra. These 1D spectra are depicted
in Fig.~\ref{fig:HgP_res} and demonstrate that the amfX2C results precisely reproduce the reference 4c spectra in both the free and
cavity-embedded cases.

The off-resonant coupling produces distinct signatures in the spectra, including the appearance of avoided crossing polaritonic branches
around all resonant frequencies in the 2D spectra, and the splitting of the L band in the 1D spectrum when the cavity is in resonance
with the B band (blue lines in Fig.~\ref{fig:HgP_res}). Previously, performing the repeated calculations necessary to construct such 2D
spectra was computationally expensive for large, heavy-element-containing molecules---particularly when an accurate description of their
electronic structure required the inclusion of relativistic effects. However, the quasi-relativistic amfX2C Hamiltonian-based QEDFT
makes these calculations accessibke thanks to its more than fivefold speed-up and its high accuracy in reproducing the 4c results.

\subsection{\label{sec:Collective} Collective coupling effects in an AuH chain}

In most cases in polaritonic chemistry, a collective-coupling situation is considered. However, in ab initio approaches to polaritonic chemistry, such as QEDFT, commonly only
a single molecule is treated quantum mechanically, while the surrounding molecules are considered
as effectively enhancing the cavity coupling parameter $g_\alpha$.~\cite{Ruggenthaler2023, horak2025analytic}
The development of more efficient ab initio methods in polaritonic chemistry makes it possible to move
beyond this approximation by explicitly treating multiple molecules in the ensemble. A study in this
direction was conducted by Sidler et al.~\cite{Sidler2020}, in which the authors considered a chain of
\ce{N2} molecules. They demonstrated that collective coupling leads to a local modification of chemical
properties on an impurity, represented by one of the \ce{N2} molecules with a perturbed bond length.

Here, we consider a similar setup: a chain of five AuH molecules, with the central molecule having an
elongated bond length and serving as the impurity. The molecules are arranged along the $x$ axis with a spacing
of \unit[20]{\AA} between them and are oriented along the $z$ axis. The ensemble molecules have an equilibrium
bond length of \unit[1.52385]{\AA}~\cite{Huber1979}, while the bond length of the impurity molecule is
\unit[1.56057]{\AA}, corresponding to the same percentage distortion as in Ref.~\citenum{Sidler2020}.
The molecules are embedded in an ideal Fabry--P{\'e}rot cavity with a single effective mode in resonance
with the excitation to the \ce{0^+} excited state at \unit[3.1676]{eV} (in the perturbed molecule, this
excitation energy is shifted to \unit[3.0911]{eV}). Due to the large separation between the molecules,
they interact only via the cavity.

\begin{figure*}
\centering
\caption{Calculated (amfX2C) absorption spectra of a chain of AuH molecules in cavities (in blue) with different coupling strengths
$g_\alpha$ (in au), compared with the spectra of an individual free molecule with equilibrium (in orange) and perturbed (in red) bond lengths.}
  \begin{subfigure}{0.32\textwidth}
    \includegraphics[width=\textwidth]{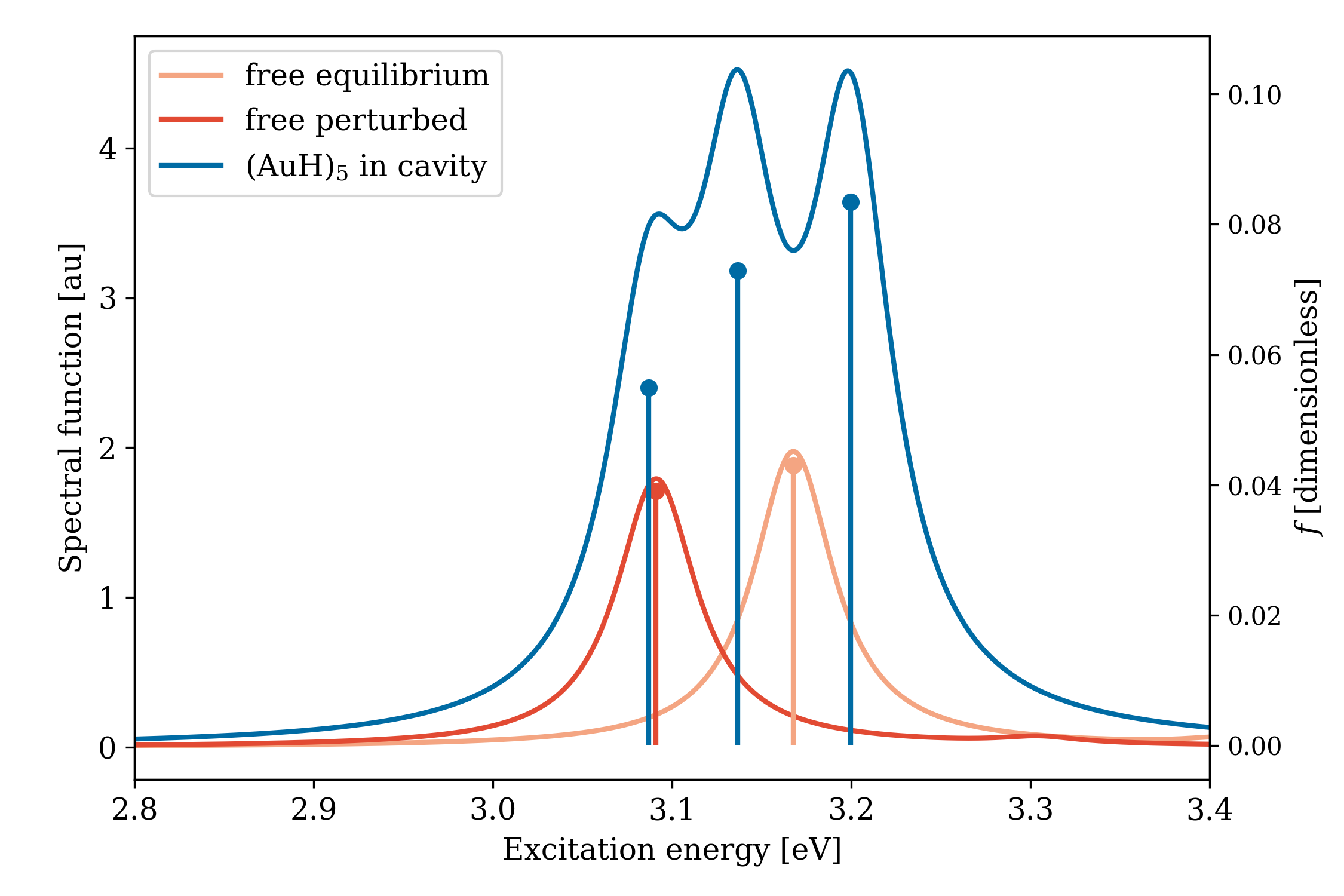}
    \caption{$g_\alpha = 0.0033$}
    \label{fig:AuH_g0033}
  \end{subfigure}
  \begin{subfigure}{0.32\textwidth}
    \includegraphics[width=\textwidth]{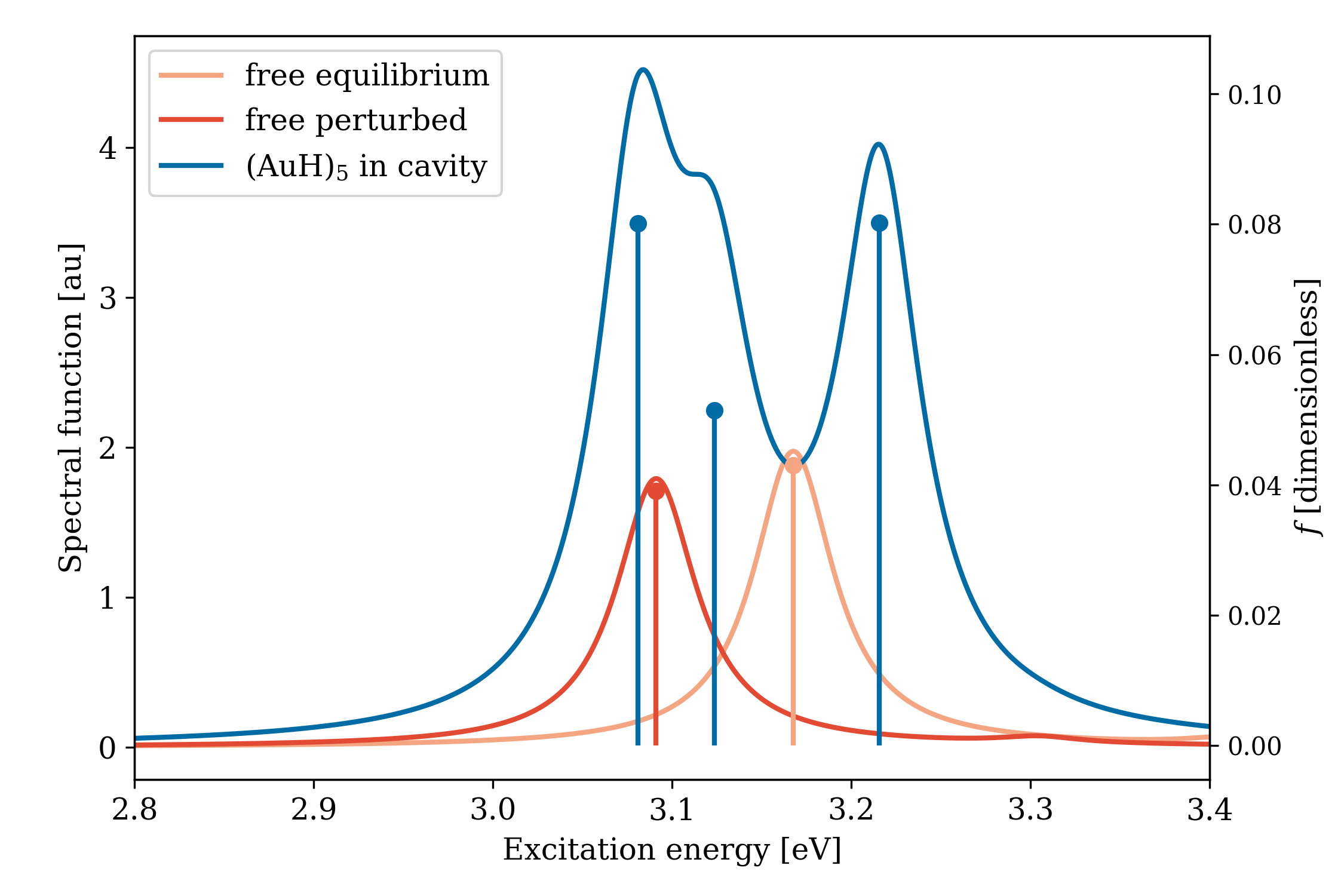}
    \caption{$g_\alpha = 0.005$}
    \label{fig:AuH_g005}
  \end{subfigure}
  \begin{subfigure}{0.32\textwidth}
    \includegraphics[width=\textwidth]{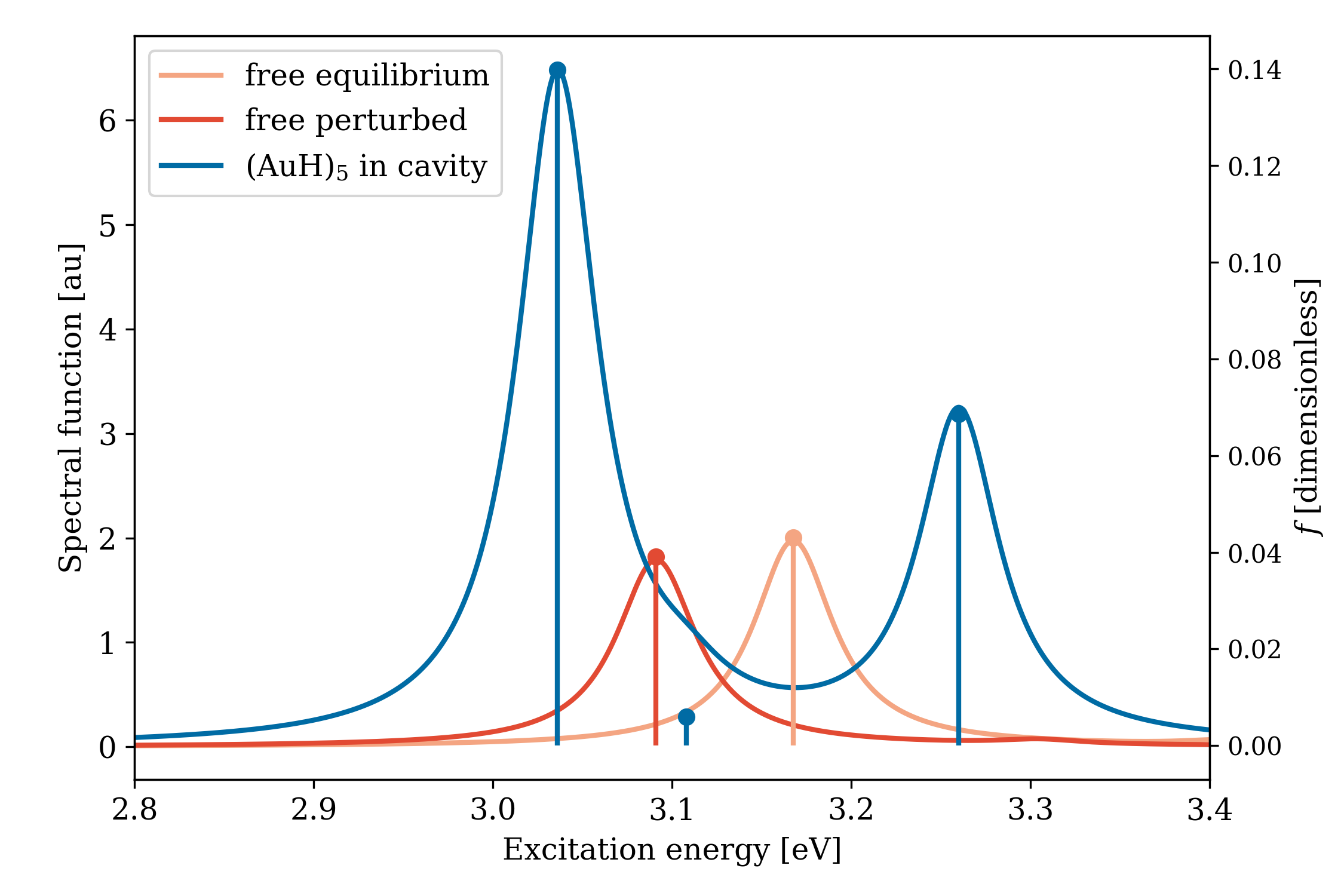}
    \caption{$g_\alpha = 0.01$}
    \label{fig:AuH_g01}
  \end{subfigure}
  \label{fig:AuH_chain}
\end{figure*}

In Fig.~\ref{fig:AuH_chain} we plot the spectra of the AuH chain for three different
coupling strengths, $g_\alpha = 0.0033$, 0.005, and 0.01, respectively, representing
three different coupling regimes. In the first regime ($g_\alpha = 0.0033$) the signals
from the equilibrium and perturbed molecules are separated and identifiable in the
spectrum. The  polaritons come only from the equilibrium-geometry molecules. In the
second regime ($g_\alpha = 0.005$), more complex effects come into play and three polaritonic
states form: lower, middle, and upper polaritons. In the third regime ($g_\alpha = 0.01$),
the Rabi splitting is significantly larger than the difference between excitation energies
of the perturbed and equilibrium molecules. The spectrum is dominated by the lower and upper
polaritons with the the middle polaritonic branch becoming dark. This is in line with the behavior of the ensemble of \ce{N2} molecules in Ref.~\citenum{Sidler2020}.

\begin{figure*}
\centering
\caption{Calculated (amfX2C) transition densities for different states in a chain of AuH molecules in a cavity
with coupling strength $g_\alpha = \unit[0.005]{au}$.}
  \begin{subfigure}{0.24\textwidth}
    \includegraphics[width=0.95\textwidth]{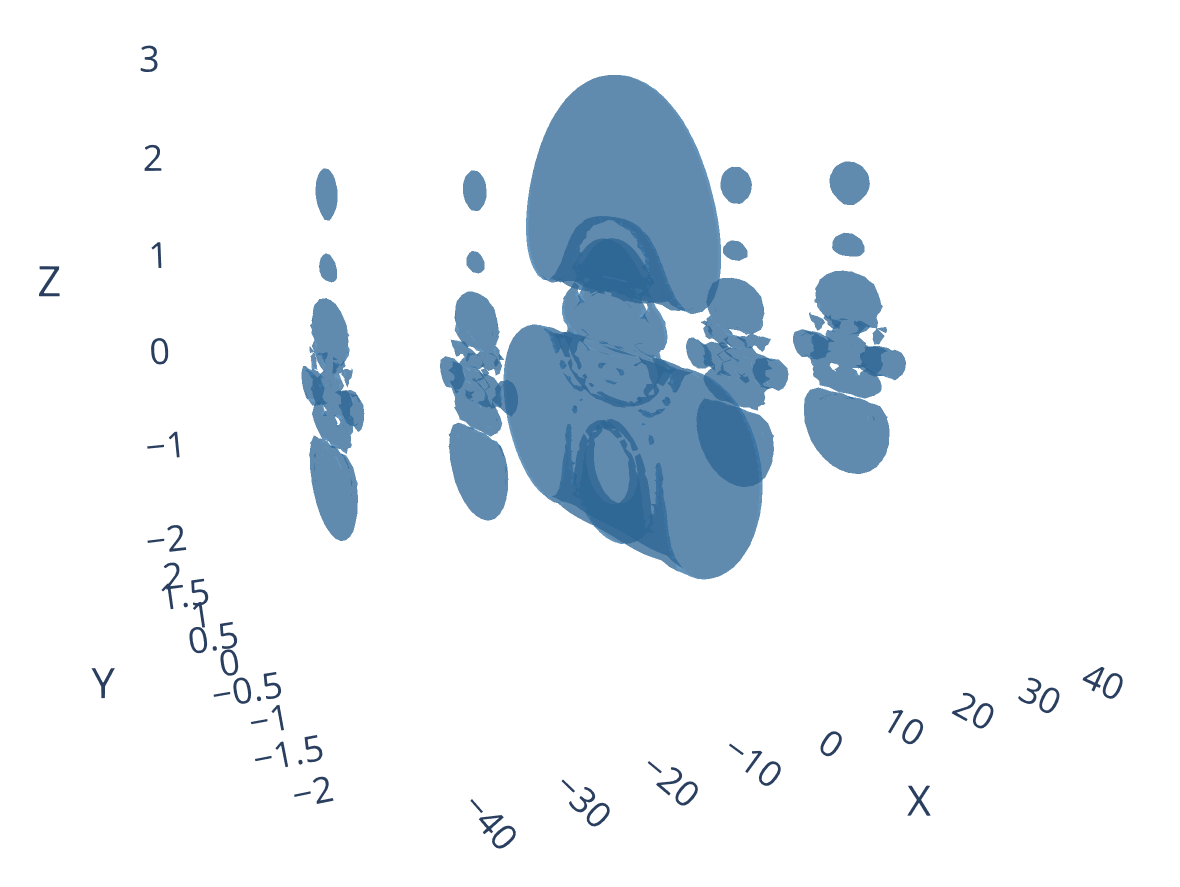}
    \caption{Lower polariton}
    \label{fig:AuH_g005_LP}
  \end{subfigure}
   \begin{subfigure}{0.24\textwidth}
    \includegraphics[width=0.95\textwidth]{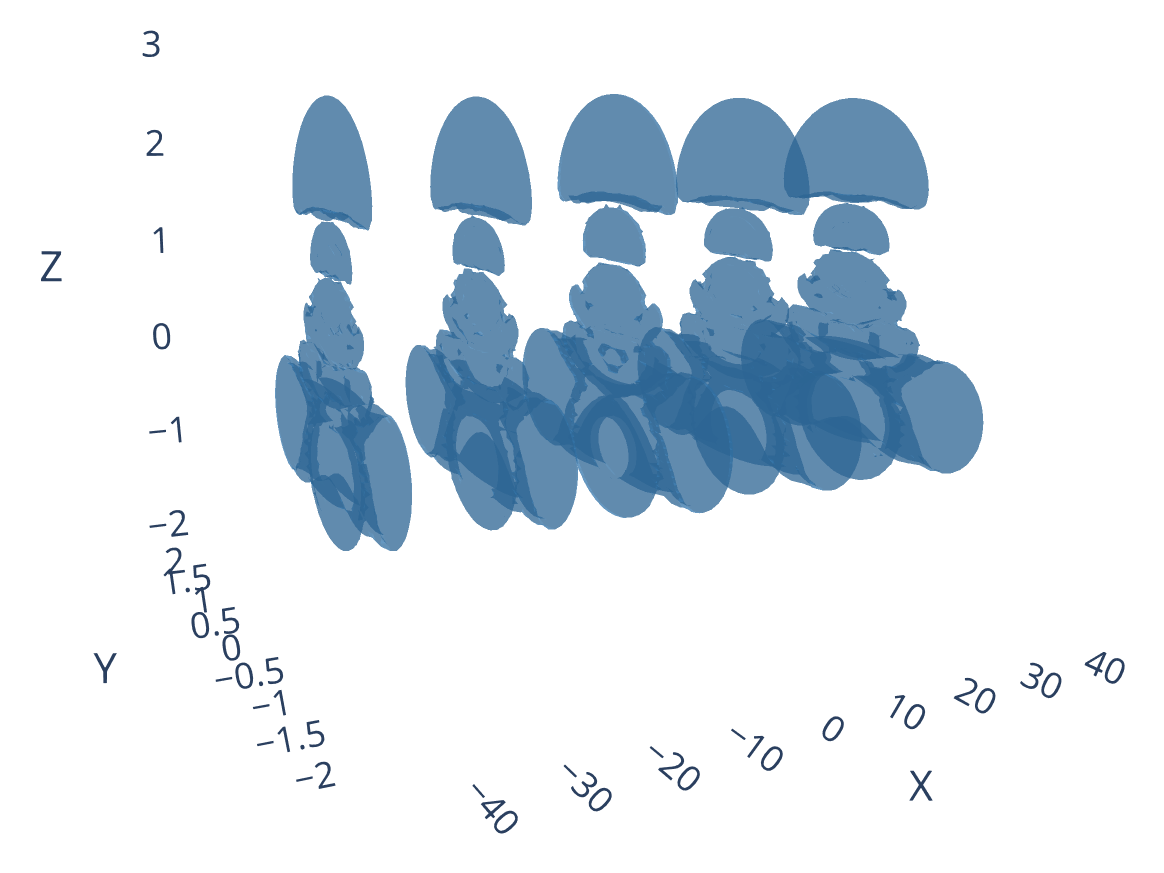}
    \caption{Middle polariton}
    \label{fig:AuH_g005_MP}
  \end{subfigure}
   \begin{subfigure}{0.24\textwidth}
    \includegraphics[width=0.95\textwidth]{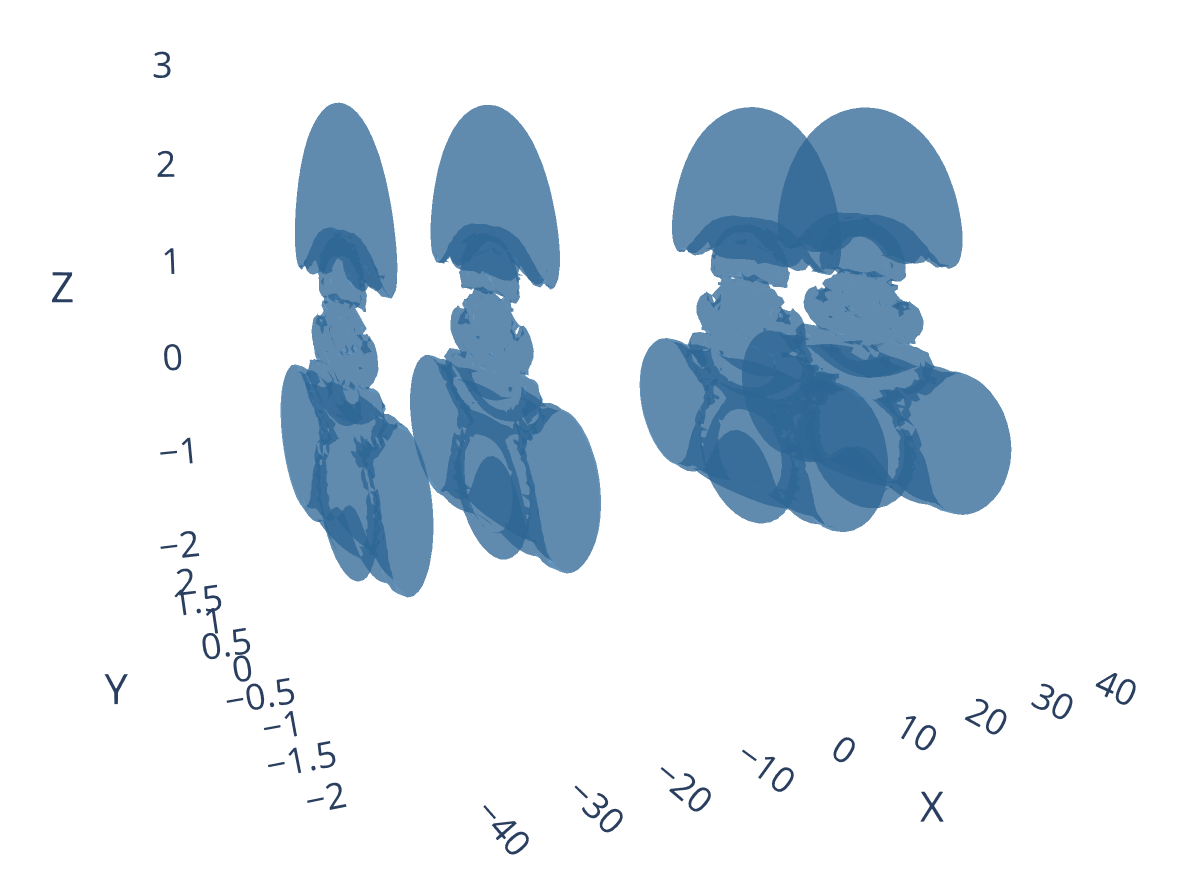}
    \caption{Dark state}
    \label{fig:AuH_g005_DS}
  \end{subfigure}
  \begin{subfigure}{0.24\textwidth}
    \includegraphics[width=0.95\textwidth]{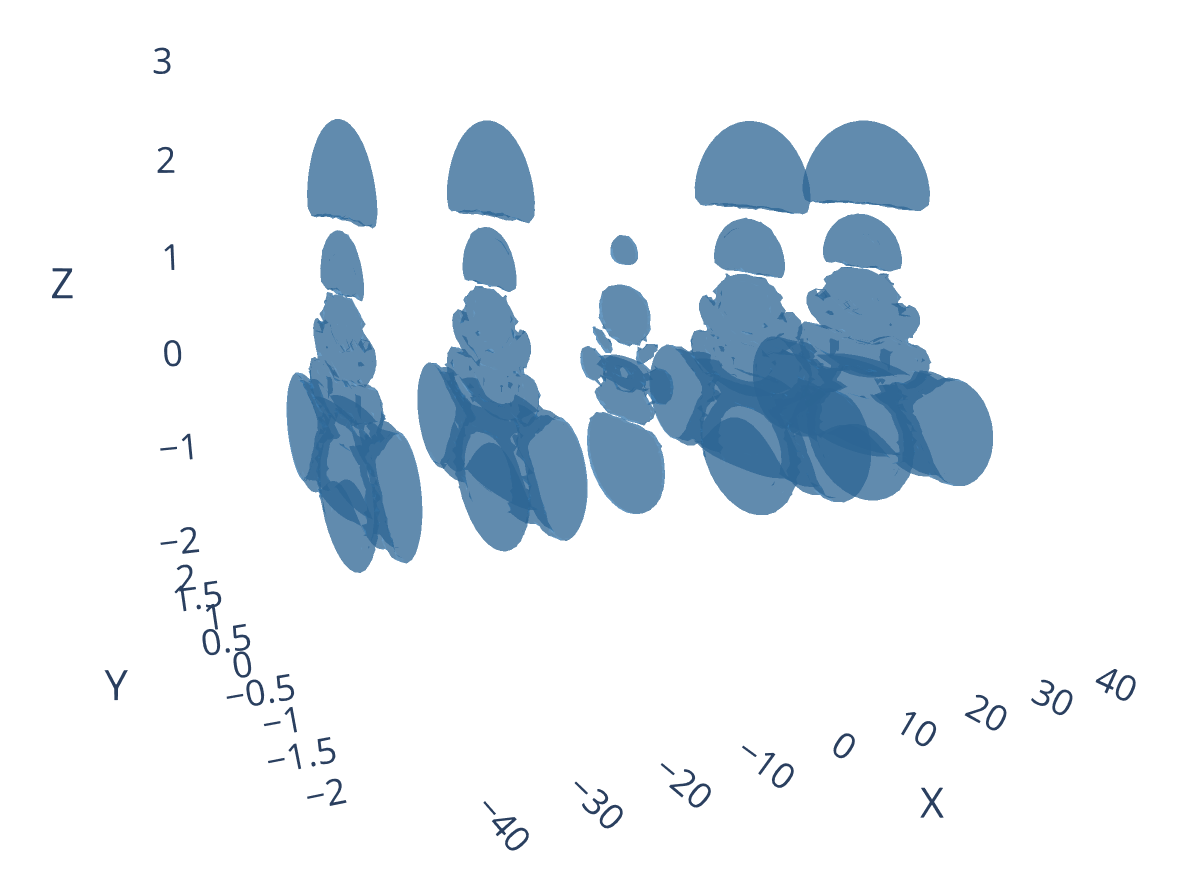}
    \caption{Upper polariton}
    \label{fig:AuH_g005_UP}
  \end{subfigure}
  \label{fig:AuH_td}
\end{figure*}

\begin{figure}
    \centering
    \includegraphics[width=0.5\linewidth]{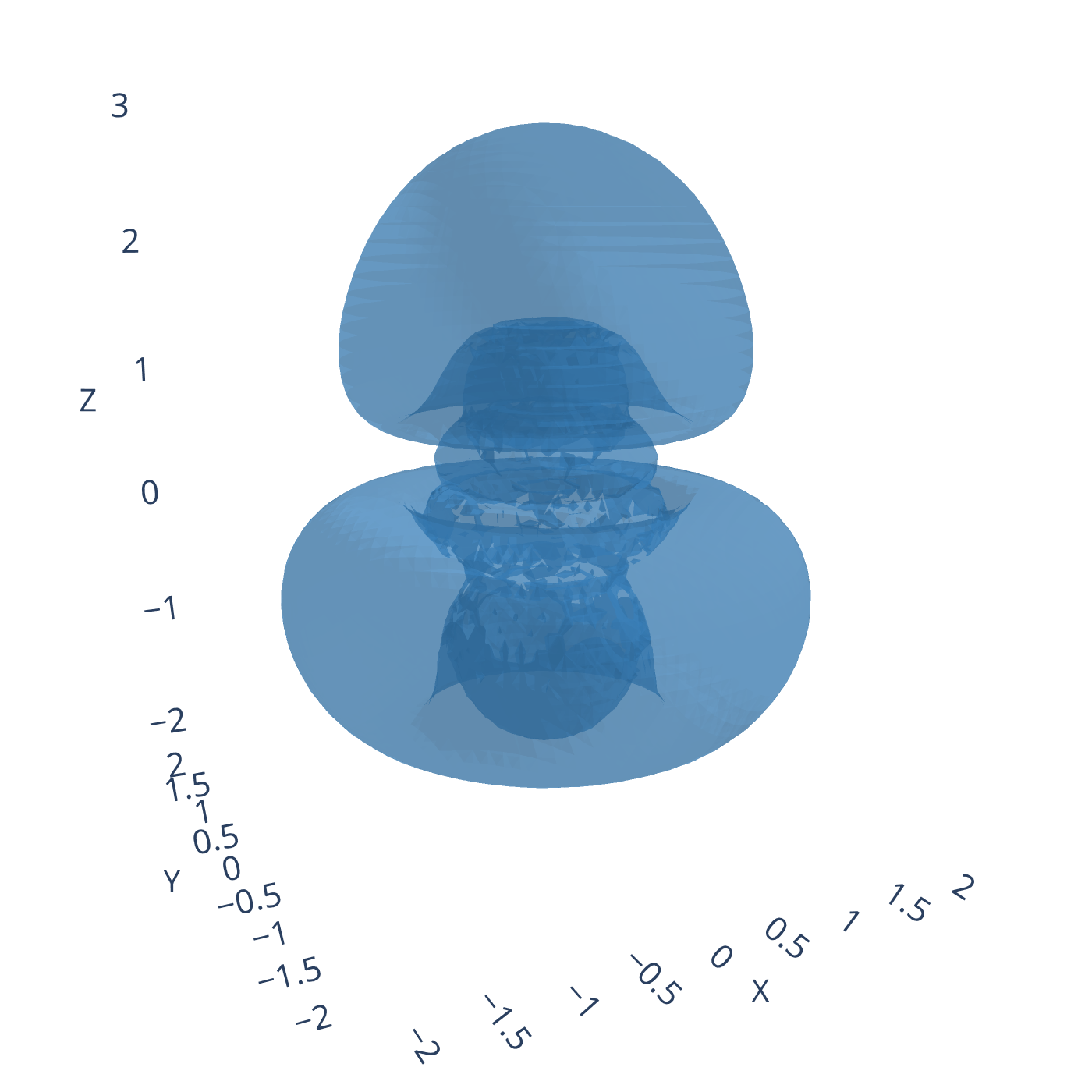}
    \caption{Calculated (amfX2C) transition density on the central AuH molecule in a five-molecule long chain corresponding to the lower polariton state in a cavity with coupling strength $g_\alpha = \unit[0.005]{au}$ (detail of Fig.~\ref{fig:AuH_g005_LP}).}
    \label{fig:AuH_td_detail}
\end{figure}

In Fig.~\ref{fig:AuH_td} we plot transition densities for the various states in the spectrum of the AuH chain
with coupling strength $g_\alpha = 0.005$ as an isosurface at value $6.0\times10^{-4}$. These states correspond
to spectral lines in Fig.~\ref{fig:AuH_g005}, namely the lower polariton at \unit[3.0809]{eV}, middle polariton
at \unit[3.1236]{eV}, a dark state at \unit[3.1670]{eV} (not seen in the spectrum), and an upper polariton at \unit[3.2158]{eV}.
Since the densities in Fig.~\ref{fig:AuH_td} are distorted due to the much larger span of the $x$ axis
we also show a detail of the transition density on the perturbed molecule for the lower polariton state
in Fig.~\ref{fig:AuH_td_detail}. These results corroborate for a case of heavy element-containing molecules
the findings of collectively-induced local effects of Ref.~\citenum{Sidler2020}. We see that the collective coupling modifies local chemical
properties at the impurity. Moreover, the unperturbed environment molecules behave identically without
boundary effects meaning that such a setup with few environment molecules can already be used to simulate
collective phenomena. Due to the all-electron nature of our calculations, a system of five AuH molecules
has more electrons as well as basis functions than the mercury porphyrin complex studied in Sec.~\ref{sec:HgP}.
This underscores the requirements for computational efficiency required for the study of collective
coupling effects in ensembles of heavy element-containing molecules. The presented results demonstrate
that amfX2C-based linear response QEDFT is ready for this task.

\section{\label{sec:Conclusions} Conclusions}

In the presented work we combined the relativistic linear response formulation of QEDFT with modern
two-component X2C Hamiltonian models, such as (extended) atomic mean-field X2C ((e)amfX2C) and molecular mean-field X2C (mmfX2C). We showed that under the conditions normally considered in the linear response regime---weak external field
limit and the dipole approximation---any dependence on time, external field, or the photon displacement
coordinate in the decoupling matrix can be approximately discarded. Therefore, it is sufficient to perform the
X2C decoupling procedure only once at the stage of the reference SCF calculation. The following
QEDFT calculation can then proceed at the two-component level. The electron--photon interaction
terms are picture change transformed using the same decoupling matrix. The two-component equations of
motion for the light--matter system can then be solved using perturbation theory (for the external field) to obtain a linear
response QEDFT equation composed of two-component quantities. The equation has the form of an eigenvalue
equation and is solved for excitation energies and transition vectors. These can be used to evaluate
electron absorption spectra as well as transition densities to provide local information about chemically
relevant changes induced by the coupling to the cavity.

The resulting X2C-based method is both efficient and accurate allowing the extension of relativistic
approach to more computationally demanding problems within polaritonic chemistry. We studied two exemplary
cases. First, for a mercury porphyrin complex we considered the molecule embedded in cavities of different
mode frequencies to construct 2D spectra. These revealed that due to the close proximity of excited states,
off-resonant coupling is significant, resulting in the appearance of several polaritonic branches and a shift
of the resonance. Moreover, due to the high density of states, an ab initio approach to this system is preferable
to a few-level model-based treatment.
Second, we considered a chain of five AuH molecules with the middle one serving as an impurity by having a non-equilibrium
bond length. Such a system is used to investigate collective coupling effects by explicitly treating an ensemble of molecules. We demonstrated that effects previously observed at the non-relativistic level of theory,
such as local modification of chemical properties due collective strong coupling, occur also for a fully relativistic treatment of quantum systems.

Relativistic quantum chemical methods based on X2C Hamiltonian models---particularly modern variants such as the amfX2C
Hamiltonian---have proven applicable to a wide range of problems in relativistic quantum chemistry. In this work, we
introduced these state-of-the-art relativistic techniques to cavity QED and polaritonic chemistry. This combination
enables computationally efficient and accurate studies of novel phenomena at the intersection of relativistic effects
and strong light--matter coupling.

\begin{acknowledgments}
We acknowledge support by the Research Council of Norway through its Centres of Excellence scheme,
project No.~262695, Research grant No.~315822, and its Mobility Grant scheme, project No.~314814.
M.R. acknowledges funding from the European Union’s Horizon 2020 research and innovation program
under the Marie Sklodowska-Curie Grant Agreement no. 945478 (SASPRO2), the Slovak Research and
Development Agency (grant no. APVV-22-0488), VEGA no. 1/0670/24 and 2/0118/25, and
the EU NextGenerationEU through the Recovery and Resilience Plan for Slovakia under the
project No.~09I05-03-V02-00034.
The calculations were performed on resources provided by Sigma2 the National Infrastructure
for High Performance Computing and Data Storage in Norway, Grant No.~NN14654K.
The Flatiron Institute is a division of the Simons Foundation. M.R. and L.K. thank Stanislav Komorovsky for valuable discussions.
\end{acknowledgments}

\section*{Data Availability Statement}

The data that support the findings of this study are available
from the corresponding author upon reasonable request.

\appendix

\bibliography{references}

\end{document}